\documentclass[
aps,superscriptaddress,
twocolumn,prl,
showpacs,preprintnumbers,nofootinbib,
amsmath,amssymb,
aps,
superscriptaddress]{revtex4-1}
\usepackage{Preamble}
\makeatletter
\AtBeginDocument{\let\includegraphics\saved@includegraphics}
\usepackage{hyperref}

\newcommand{\bea}{\begin{eqnarray}}
\newcommand{\eea}{\end{eqnarray}}

\begin{document}

\title{
Reducing the irreducible:
the charged black hole bomb in a moving cavity}


\author{Nicolas Sanchis-Gual}
\affiliation{Departamento de Astronomía y Astrofísica, Universitat de València, Dr.\ Moliner 50, 46100, Burjassot (Valencia), Spain}

\author{Alejandro Belchí}
\affiliation{Departamento de Astronomía y Astrofísica, Universitat de València, 
Dr.\ Moliner 50, 46100, Burjassot (Valencia), Spain}

\author{Carlos Herdeiro}
\affiliation{Departamento de Matemática da Universidade de Aveiro 
and Centre for Research and Development in Mathematics and Applications (CIDMA)
Campus de Santiago, 3810-193 Aveiro, Portugal}

\author{José A. Font}
\affiliation{Departamento de Astronomía y Astrofísica, Universitat de València, 
Dr.\ Moliner 50, 46100, Burjassot (Valencia), Spain}
\affiliation{Observatori Astron\`{o}mic, Universitat de Val\`{e}ncia,
C/ Catedr\'{a}tico Jos\'{e} Beltr\'{a}n 2, 46980, Paterna (Val\`{e}ncia), Spain}

\begin{abstract}

We revisit the charged black hole bomb by numerically solving the fully non-linear Einstein–Maxwell–(charged, complex) Klein–Gordon system with a moving mirror. By dynamically varying the cavity size, we find that the system evolves toward new hairy black hole equilibria. Expanding the mirror radius enhances superradiant extraction, increasing both the scalar field charge and the black hole’s irreducible mass. Remarkably, on the other hand, shrinking the cavity size has the opposite effect: the black hole is able to reduce its irreducible mass as more charge than energy flows back from the field, without violating charge conservation or energy conditions. As a consistency check, in the limit of a vanishing cavity, we find that the system returns to the original Reissner–Nordström configuration. We discuss the implications of these findings for black hole thermodynamics in confined configurations where superradiant modes exist and the limitations of this setup, particularly in relation to Hawking's black hole area theorem.

\end{abstract}

\maketitle

\textit{\textbf{Introduction}.} Black holes (BHs) are extraordinary laboratories to test the rationale behind the physical laws governing our Universe due to their extreme conditions and properties. Singularities and horizons represent a challenge to the understanding of classical and quantum gravity theories alike, exposing the limits of General Relativity and emphasizing the need for a consistent theory of quantum gravity. These regions highlight profound conceptual issues, such as the true nature of spacetime near singularities and the fate of information behind event horizons~\cite{penrose1965gravitational,hawking1971gravitational,hawking1976breakdown,preskill1992black}.

One of the most striking discoveries in the study of BHs is the realization that they behave like thermodynamic systems: they possess temperature, emit radiation, and obey laws that closely resemble the classical laws of thermodynamics~\cite{bekenstein1974quantum,hawking1975particle}. Central to this analogy is the area theorem, or Hawking's area law, which states that the area of a BH's event horizon, and thus its irreducible mass, cannot decrease in any classical process, provided that the null energy condition holds~\cite{hawking1971gravitational}. This theorem supports the interpretation of BH area as a measure of entropy and is fundamental to our current understanding of BH thermodynamics.

While quantum effects such as Hawking radiation provide mechanisms for BH mass loss, these operate on a semiclassical framework and are not constrained by the classical area theorem. Here, however, we focus entirely on classical dynamics examining the nonlinear evolution of BHs in the Einstein-Maxwell-Klein-Gordon (EMKG) system, where a charged scalar field interacts with a Reissner-Nordström (RN) BH confined within a  cavity~\cite{Degollado:2013bha,Degollado:2013eqa,Sanchis-Gual:2015lje}.

The confinement of scalar fields around BHs, first envisioned in the so-called BH bomb mechanism by Press and Teukolsky~\cite{Press:1972zz}, provides a setting for superradiant instabilities~\cite{brito2020superradiance}: charged or rotating BHs can amplify incident waves under suitable conditions~\cite{east2014black}, leading to exponential growth of the field and extraction of energy and charge from the BH. Past studies of such instabilities have often assumed fixed boundary conditions, such as a mirror placed at a constant radius, or a potential well due to a mass term, and demonstrated that the system dynamically evolves towards equilibrium configurations featuring scalar or vector “hair”~\cite{dolan2015stability,Sanchis-Gual:2015lje,Bosch:2016vcp,Sanchis-Gual:2016tcm, baake2016superradiance,east2017superradiant1}.

In this work, we go beyond the standard setup and investigate the dynamical response of the system when the mirror's location is no longer fixed. Specifically, we first allow the system to evolve towards a stationary hairy BH with the mirror placed at a fixed radius. Once equilibrium is reached, we adiabatically vary the position of the mirror, treating it as a movable boundary. While not straightforwardly tied to any specific known astrophysical process, it provides valuable insight into the behavior of confined BHs and their interaction with bosonic fields. Thus, it serves as a theoretical probe to explore the stability and thermodynamic response of the system under time-dependent confinement. This approach could serve as a proxy for more general confining mechanisms, including effective potentials or asymptotic Anti-de-Sitter (AdS)-like boundaries.

We find that if the mirror is pushed outward, the system resumes the superradiant extraction of charge and energy, eventually relaxing to a new equilibrium configuration corresponding to the new cavity size. The final state features a larger apparent horizon mass and a higher irreducible mass, fully consistent with the classical area theorem.

When the mirror has instead a inward motion, the opposite behavior is observed: the scalar field returns some of its charge to the BH, and the system relaxes to the new stationary configuration fitting to the reduced cavity size. Moreover, we observe that the BH’s irreducible mass decreases during this process. If interpreted thermodynamically, this would indicate a decrease in the entropy of the BH, a seemingly paradoxical result in a purely classical system. We argue that actually this does not constitute a violation of the area theorem {\it per se}, since the scalar field and cavity outside the BH act as a classical reservoir and the full system (BH + environment) may still obey a generalized second law. Nonetheless, this behavior raises subtle questions about the status of the area theorem in confined, dynamical, superradiantly unstable systems and in the presence of non-trivial boundary conditions, $e.g$ in asymptotically AdS spacetimes.

Our main analysis is performed in spherical symmetry using numerical relativity techniques adapted to the EMKG system. This ensures that no energy or charge escapes the system and gravitational and electromagnetic radiation are absent by symmetry. 
These results, while limited to spherical symmetry, open the door to exploring BH thermodynamics in more complex, dynamical, and interacting settings, where traditional assumptions may require refinement. 

{\bf {\em Framework.}} We consider the EMKG system, governed by the action $\mathcal{S}=\int d^4x \sqrt{-g}\mathcal{L}$, where the Lagrangian density is given by: \begin{equation} \mathcal{L}=\frac{R-F_{\alpha\beta}F^{\alpha\beta}}{16\pi}-\frac{1}{2}D_\alpha \Phi (D^\alpha\Phi)^*-\frac{\mu^2}{2}|\Phi|^2 \ , \label{model} \end{equation}
with $R$ being the Ricci scalar, and $F_{\alpha \beta}\equiv \nabla_{\alpha}A_{\beta} - \nabla_{\beta}A_{\alpha}$ the usual electromagnetic field strength tensor. Here, $A_{\alpha}$ is the electromagnetic potential, and $D_\alpha\equiv \nabla_\alpha -iqA_\alpha$ denotes the gauge covariant derivative. The scalar field carries charge $q$ and mass $\mu$. Here, we will consider a massless scalar field $\mu=0$ for simplicity, although we have also evolved a massive field and the results hold. We adopt natural units by setting Newton's constant, the speed of light, and $4\pi\epsilon_0$ to one.

Following~\cite{Sanchis-Gual:2015lje}, we employ a generalized Baumgarte-Shapiro-Shibata-Nakamura (BSSN) formulation~\cite{Baumgarte98,Shibata95}, adapted for spherical symmetry~\cite{Brown:2009,Alcubierre:2010is,Montero:2012yr}, to numerically evolve the EMKG system, using the code developed in~\cite{Sanchis-Gual:2015bh,Sanchis-Gual:2015sms}. The 3+1 decomposition of the spacetime metric takes the form $ds^2=-(\alpha^2+\beta^r \beta_r)dt^2+2\beta_r dtdr+e^{4\chi}\left[a\, dr^2+ b\, r^2 d\Omega_2\right]$, where the lapse $\alpha$, the radial component of the shift vector $\beta^r$, and the spatial metric functions $\chi,a,b$ are functions of $t$ and $r$. The electric field is defined as $E^\mu=F^{\mu\nu} n_\nu$ and possesses only a radial component, while the magnetic field $B^\mu=\star F^{\mu\nu} n_\nu$ vanishes. The vector $n^\mu$ denotes the 4-velocity of the Eulerian observer~\cite{Torres:2014fga}. 

At the mirror location $r=r_{m}$ and beyond, the scalar field $\Phi$ is set to vanish (Dirichlet boundary conditions). We impose parity boundary conditions on the scalar field at the origin (puncture). The EMKG system admits the RN BH as a solution, with ADM mass $M$ and charge $Q$, and a vanishing scalar field. We choose $M=1$ and $Q=0.9$ for our simulations, but the mass merely sets the natural scale of the problem. Perturbing this background with a spherically symmetric scalar Gaussian wave can lead to a superradiant instability, provided that $\omega<\omega_{\rm c} \equiv q\phi_{\rm H}$, where $\omega$ is the oscillation frequency of the field and $\phi_{\rm H}$ is the electric potential at the event horizon.

We track the irreducible mass~\cite{Christodoulou:1970wf} of the BH, which is computed from the apparent horizon (AH) area $A_{\rm{AH}}$ at each time slice as $M_{\rm{irr}} = \sqrt{{A_{\rm{AH}}}/(16\pi)}$. We observe that the extraction of energy and charge from the BH due to superradiant instability is consistent with the second law of thermodynamics and therefore the area law. However, during the contraction of the cavity, we observe that the irreducible mass decreases.  

Additionally, we monitor the electric charge associated with the BH and with the scalar field. On one hand, the scalar field charge is obtained via the spatial volume integral to compute the charge transferred from the BH to the scalar field \begin{equation}
Q_{\rm{SF}}=\int^{r_{m}}_{r_{\rm{AH}}}j^{\rm{SF}}_e dV, \label{eq:ESF} 
\end{equation} 
where $r_{\rm AH}$ is the horizon radius and $j^{\rm{SF}}_e$ is the projection of the scalar field’s charge density along the normal to constant-$t$ hypersurfaces~\cite{Alcubierre08a,Torres:2014fga}. On the other hand, the BH charge, $Q_{\rm{BH}}$, is evaluated at the apparent horizon as~\cite{Torres:2014fga} \begin{equation} Q_{\rm{BH}}=\left(r^{2}e^{6\chi}\sqrt{ab^{2}}E^{r}\right)\big|_{\rm AH} \ . 
\end{equation}

\textit{\textbf{Results: increasing the cavity (case 1)}.} We explore the effect of increasing the size of the cavity by increasing the radius of the mirror $r_{m}$, treating it as a movable piston rather than a fixed boundary. The trajectory of the mirror follows:
\begin{eqnarray}
r_{m}(t)&=&r_{m}^{\rm init}= r_{m}(t_{\rm init}),\quad t<t_{\rm init}\nonumber\\
r_{m}(t)&=&r_{m}^{\rm init}+v_{m}\,(t-t_{\rm init}),\quad t_{\rm init}<t<t_{\rm final}\nonumber\\
r_{m}(t)&=&r^{\rm final}_{ m}= r_{m}(t_{\rm final}),\quad t>t_{\rm final}\nonumber,
\end{eqnarray}
where $t_{\rm init}$ and $t_{\rm final}$ are the initial and final times of the mirror's motion, respectively, and $v_{m}$ is the velocity of the mirror, satisfying $v_{m} < 1$.

This setup is physically motivated: as superradiant scalar modes grow, they exert pressure on the mirror, naturally pushing it outward. This idea resembles the spirit of the original Press and Teukolsky BH bomb, where the mirror is fixed, but the growing pressure from the scalar field is eventually expected to destroy the fixed mirror. In our scenario, the system offers a variation on that classic setup: a BH bomb within a moving mirror (piston). Then, the motion of the mirror is natural. In fact, the BH-mirror system could well be isolated. Throughout this process, the entropy of the BH increases, and so does the total entropy of the system, assuming no entropy is attributed to the scalar environment. 

\begin{figure}[t!]
\begin{center}
\includegraphics[width=0.48\textwidth]{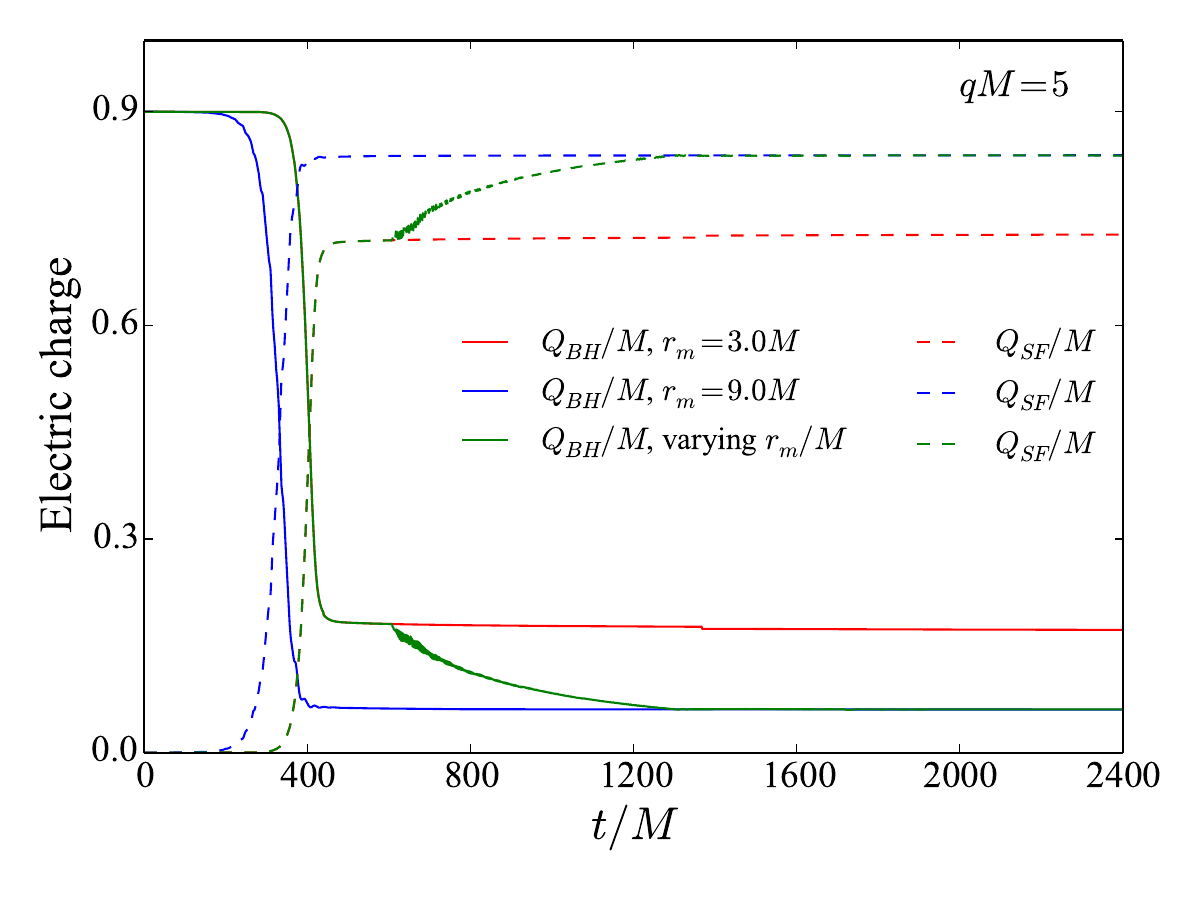}\\
\includegraphics[width=0.46\textwidth]{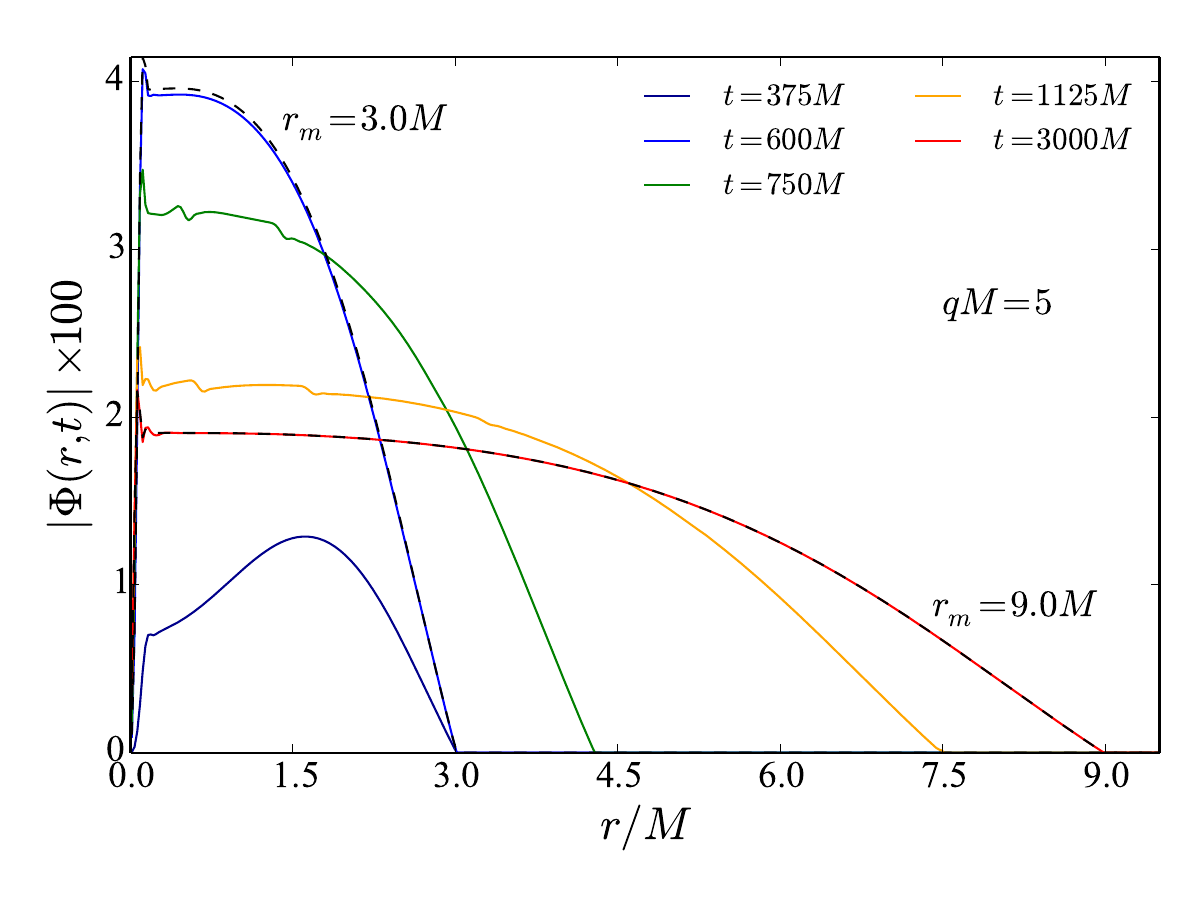}\vspace{-0.5cm}\\
\caption{(Top panel) Time evolution of the BH and scalar field charges $Q_{\rm BH}$ and $Q_{\rm SF}$ for different fixed values of the mirror radius $r_{m}/M=\lbrace3, 9\rbrace$ (red and blue curves) and a moving mirror (green curves). (Bottom panel) Radial profile of the scalar field magnitude at different times for the increasing mirror radius case with $qM=5$.}
\label{fig1}
\end{center}
\end{figure}

We proceed as follows. First, we evolve the system with a fixed mirror at $r^{\rm init}_m = 3M$ until it passes the saturation point and reaches an equilibrium state. Then, starting at $t_{\rm init} = 600M$ and continuing until $t_{\rm final} = 1300M$, we gradually move the mirror outward to $r_{m} = 9M$, with $v_{m}\sim0.0085$. The top panel of Fig.~\ref{fig1} shows the evolution of the scalar field charge $Q_{\rm SF}$ and the BH charge $Q_{\rm BH}$ for three cases: a fixed mirror at $r_{m} = 3M$ (red solid and dashed lines), a fixed mirror at $r_{m} = 9M$ (blue solid and dashed lines), and a moving mirror with $qM = 5$ (green solid and dashed lines). The bottom panel displays the radial profile of the scalar field amplitude $|\Phi(t,r)|$ at various times for the moving mirror case, with the corresponding profiles for the two fixed mirror setups shown as dashed black lines.

As the mirror moves outward, the scalar field relaxes into a new hairy BH configuration corresponding to the final mirror radius, as if the system had evolved from the start with $r_{m} = 9M$. During this transition, charge and energy extraction from the BH resumes, leading to an increase in the apparent horizon area and mass. Larger values of $r_{m}$ result in more efficient extraction of electric charge and a greater final apparent horizon mass. In the limit $r_{m} \rightarrow \infty$, the BH approaches a Schwarzschild configuration with $M_{\rm irr} \simeq M$ (see Appendix A), while all the charge $Q/M$ is stored in the surrounding scalar field cloud outside the horizon, as seen in~\cite{Bosch:2016vcp}.

\begin{figure}[t!]
\begin{center}
\includegraphics[width=0.48\textwidth]{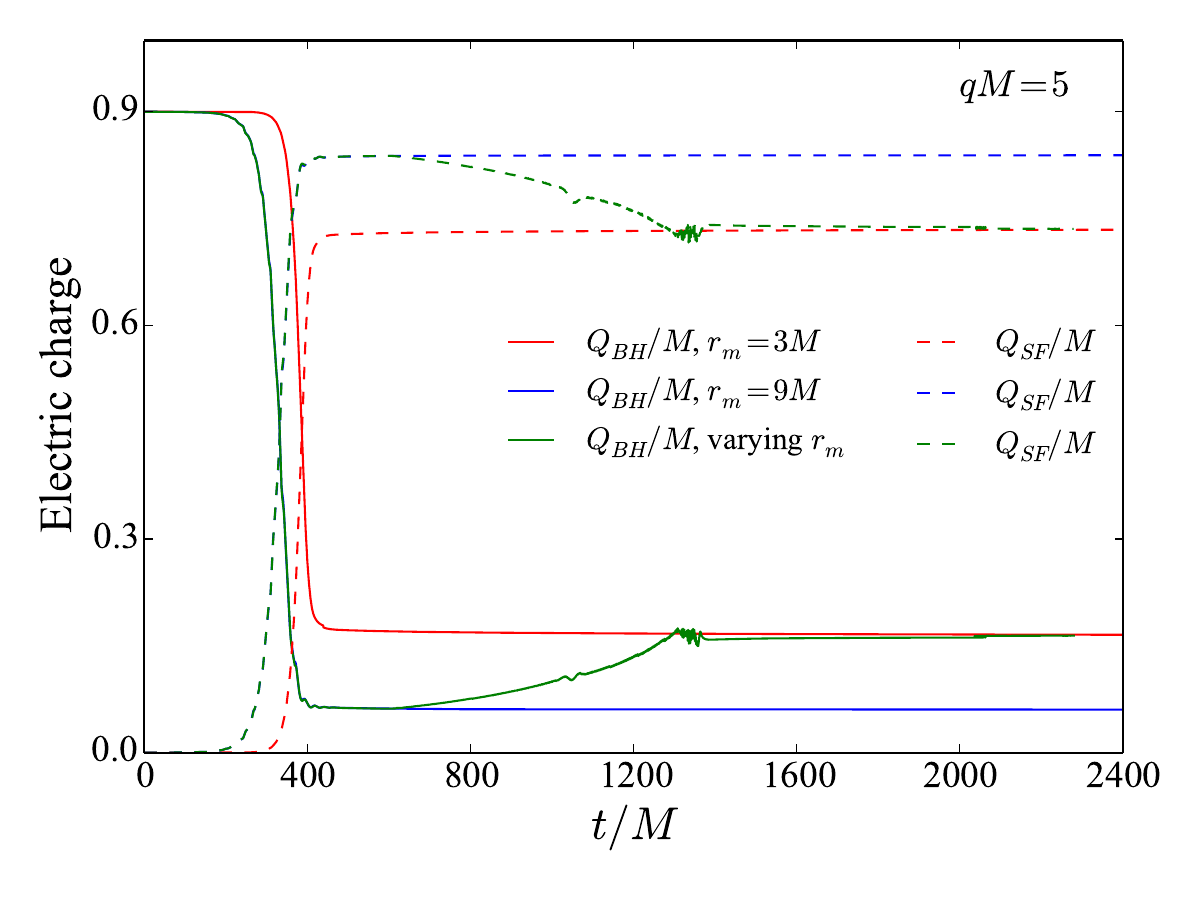}\vspace{-0.5cm}
\caption{Time evolution of the BH and scalar field charges $Q_{\rm BH}$ and $Q_{\rm SF}$ for the shrinking mirror radius case with $qM=5$.}
\label{fig2}
\end{center}
\end{figure}

However, the timescale of the superradiant instability increases proportionally with the mirror radius, following $T \sim r_{m}$, as shown in appendix~A
. This scaling arises because larger cavities support scalar field modes with longer wavelengths and lower frequencies, which grow more slowly. As a result, in the limit $r_{m} \rightarrow \infty$, the instability becomes arbitrarily slow, and an infinite amount of time would be required to fully deplete the RN BH.

\begin{figure}[t!]
\begin{center}
\includegraphics[width=0.48\textwidth]{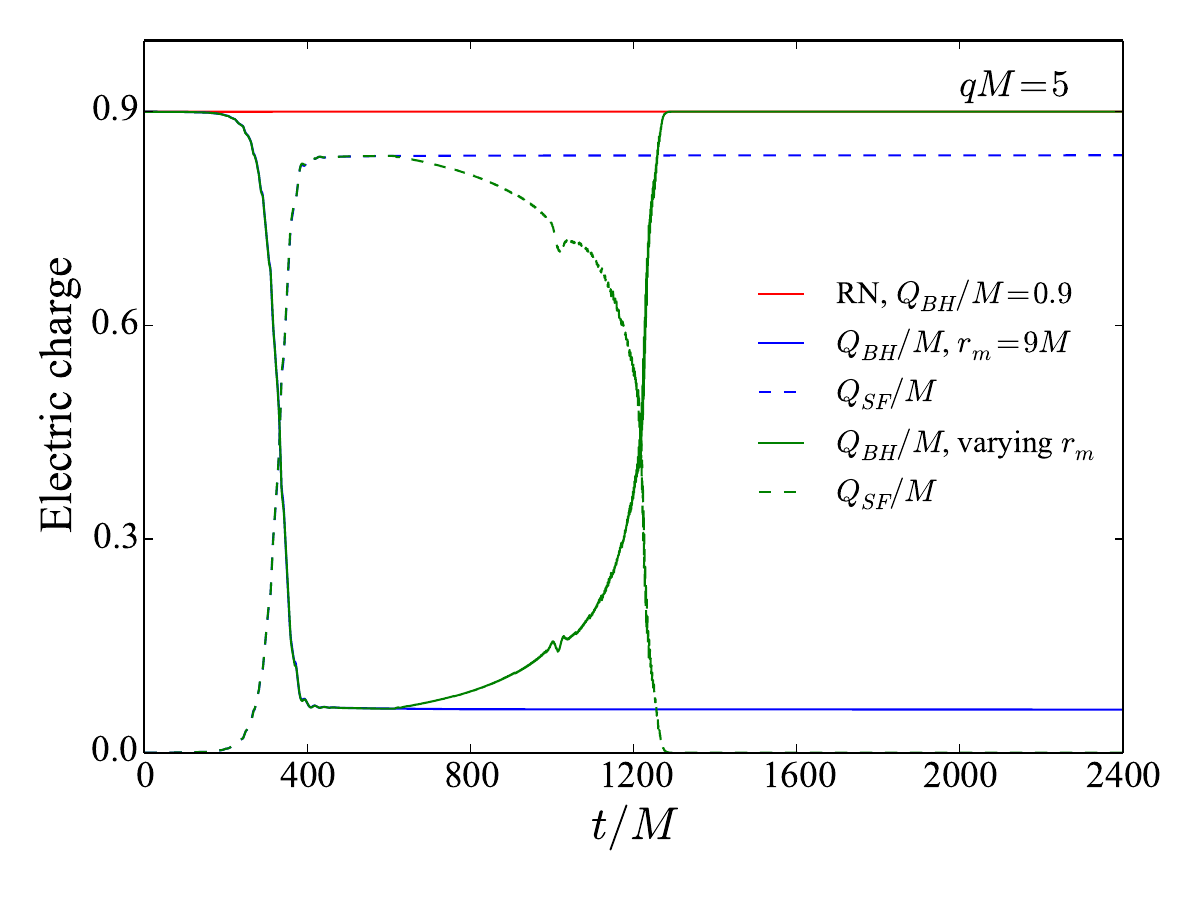}\\
\includegraphics[width=0.48\textwidth]{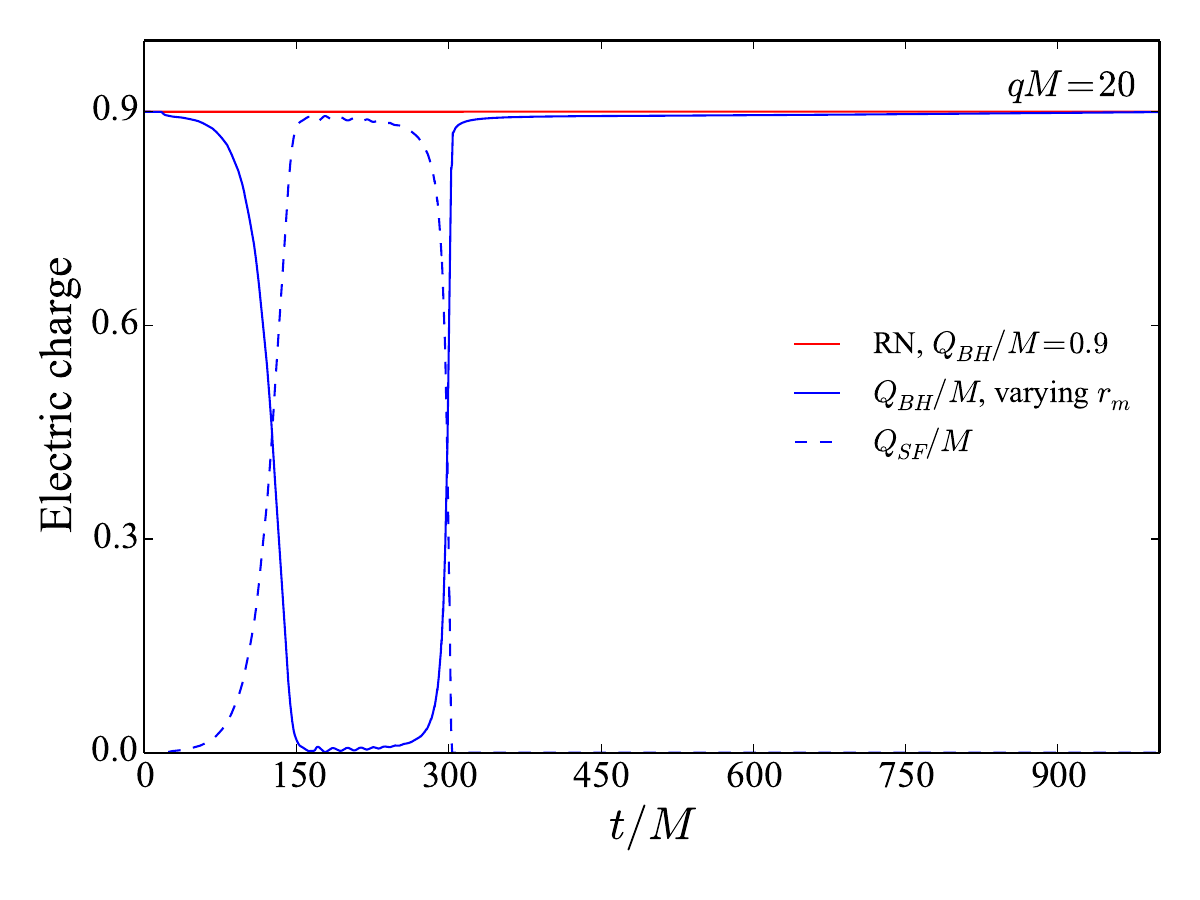}\\
\includegraphics[width=0.48\textwidth]{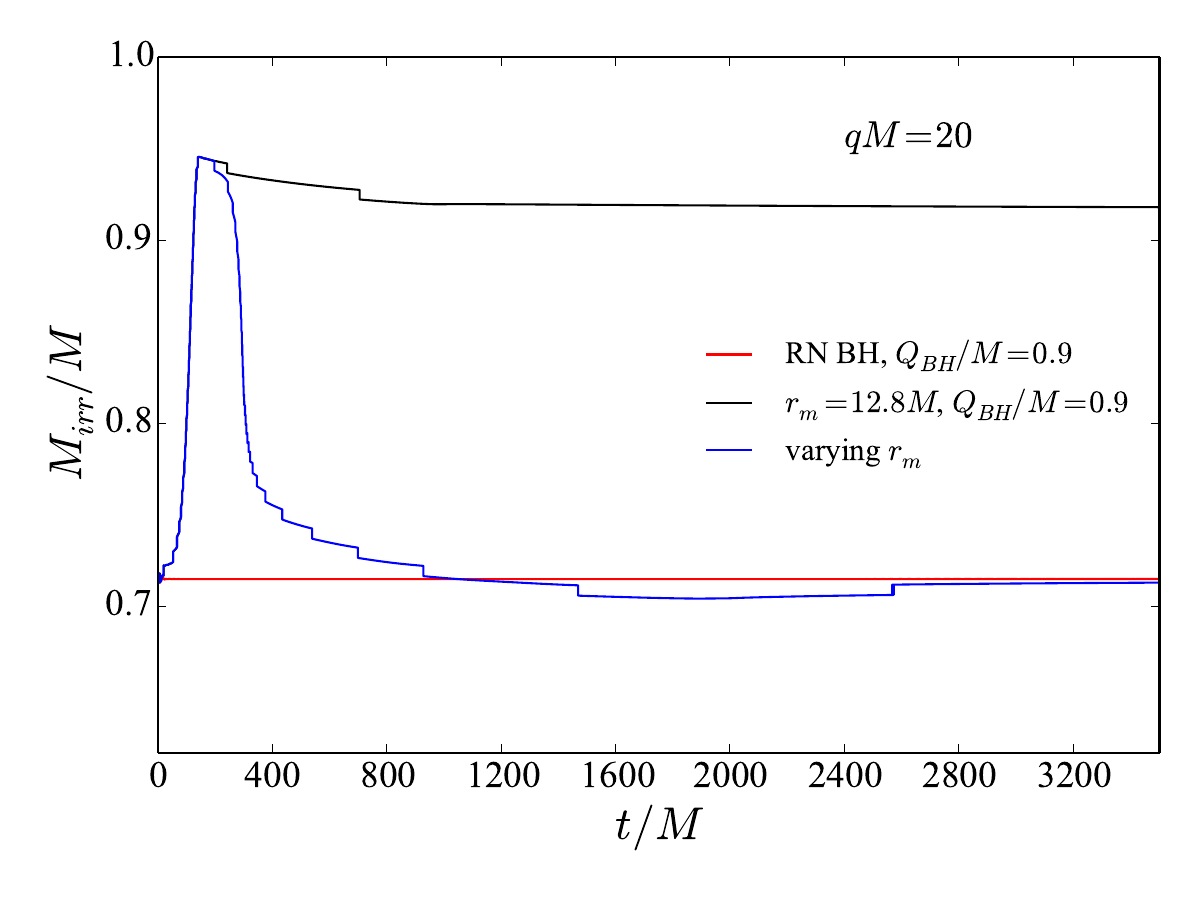}\vspace{-0.5cm}\\
\caption{Time evolution of the BH and scalar field charges $Q_{\rm BH}$ and $Q_{\rm SF}$ for $qM=5$ (top panel) and $qM=20$ (middle panel), considering a shrinking mirror reaching $r_{m}=0$. (Bottom panel) Time evolution of the apparent horizon mass for $qM=20$.}
\label{fig3}
\end{center}
\end{figure}

\textit{\textbf{Decreasing the mirror radius (case 2)}.} We now turn our attention to the complementary scenario in which the cavity size is reduced. We model the inward motion of the mirror as being driven by an external agent that adjusts the cavity radius adiabatically. This approach can be seen as analogous to a quasi-static process in thermodynamics and allows us to study how the system transitions between equilibrium states with smaller mirror radii. In this case, we observe an inverse but qualitatively analogous outcome: as the mirror radius decreases, the system evolves toward a new equilibrium configuration characterized by a lower scalar field charge. Consequently, the scalar field is compelled to shed charge, even as its energy increases.

We investigate this behavior by evolving two configurations with different scalar field charge couplings, $qM = 5$ and $qM = 20$. In Fig.~\ref{fig2}, we show the $qM = 5$ case, illustrating the reverse evolution compared to the expanding cavity scenario: the mirror radius is decreased from $r_{m} = 9M$ to $r_{m} = 3M$, following the same time profile as in case 1, with $v_{m}\sim-0.0085$. As the charge of the BH increases, $\phi_{\rm H}$ and therefore $\omega_{\rm c}$ also increase, but the frequency of the scalar field matches the critical frequency at all times.

To further test the system, we consider a more extreme scenario in which the mirror radius is reduced all the way to $r_{m} = 0$. This is actually equivalent to reduce the mirror up the apparent horizon radius. We evolve two configurations: ($r_{m}^{\rm initial} = 9M,\, qM = 5$) and ($r_{m}^{\rm initial} = 12.8M,\, qM = 20$). In the top panel of Fig.~\ref{fig3}, we show the evolution of $Q_{\rm SF}$ and $Q_{\rm BH}$ for the case with $qM = 5$, with the mirror's motion starting at $t_{\rm init}=600$ and $v_m=-0.01286$. The scalar field is gradually reabsorbed into the BH until $t \sim 1300M$, at which point it completely disappears inside the horizon. Throughout the evolution, total electric charge is conserved, and the BH charge asymptotically returns to its initial value of $Q/M = 0.9$.

We repeat the experiment for the case $qM = 20$, reducing the mirror radius from $r_{m} = 12.8M$ to $r_{m} = 0$. Fig.~\ref{fig3} (middle panel) presents the results for the electric charges with the mirror motion beginning at $t = 150M$ and reaching $r_{m} = 0$ at $t = 310M$, with $v_m=-0.08$. 
Independently of the mirror motion, we consistently observe the complete dissipation of the scalar field energy and the full return of electric charge to the BH.\footnote{A convergence test and a comparison with three-dimensional simulations can be found in Appendix~B and~C respectively.}

Shrinking the cavity leads to a striking and unexpected result. During the superradiant phase, the apparent horizon mass $M_{\rm irr}$ can increase significantly, particularly for rapidly growing modes associated with large values of $q/M$ and short instability timescales. For instance, in the case of $qM = 20$ and a fixed mirror at $r_{m} = 12.8M$, the system evolves to a hairy BH with $M_{\rm irr} \approx 0.92$, as shown in the bottom panel of Fig.~\ref{fig3}. If the charge is subsequently returned to the BH, it must accommodate a larger charge-to-mass ratio. This is achieved by reducing its irreducible mass, ultimately restoring the original RN solution with the same initial mass and charge. The blue curve in the bottom panel of Fig.~\ref{fig3} illustrates the evolution of $M_{\rm irr}$. The apparent horizon mass decreases rapidly and asymptotically approaches its initial value, shown by the red curve corresponding to a bald RN BH, with a dramatic decrease of $M_{\rm irr}$. 

Imposing boundary conditions to the field near the apparent horizon may not be physically realizable if we consider a material mirror. Such a mirror could require exotic matter that violates certain assumptions of the area theorem, raising doubt on whether a decrease in 
$M_{\rm irr}$ could occur, especially if the mirror were to fall into the black hole. We emphasize, however, that \textit{any} reduction of the mirror radius reduces $M_{\rm irr}$, even far away from the horizon, where the physical nature of the material composing the mirror does not raise major concerns (see Appendix D).

\textit{\textbf{Discussion}}  The observation that $M_{\rm irr}$ can decrease for a BH in a shrinking cavity, while the total energy and charge are conserved (see Appendix E for a discussion on energy balance), opens questions about the scope of the area theorem in dynamical but classically consistent settings. 

This work thus invites further exploration of BH thermodynamics in non-asymptotically flat, dynamical spacetimes, especially beyond spherical symmetry, where entropy bounds and energy conditions may behave in less constrained ways, or even in  analogues of black holes~\cite{torres2017rotational} or optics. Future investigations could explore these effects in rotating spacetimes, or by coupling to additional classical fields, to assess the robustness and generality of these phenomena. Finally, although the evolution is entirely classical, it may be probing a regime where quantum backreaction is essential to preserve fundamental consistency with semiclassical expectations. Specifically Schwinger pair production~\cite{schwinger1951gauge} may become non-negligible as the mirror is pushed inwards and the scalar field is compressed, as supported by the analysis in Appendix F.

\bigskip

\textit{\textbf{Acknowledgements}}. We thank Juan Calderón Bustillo, Adrián Del Río, Miguel Ángel Sanchis-Lozano, Jorge Castelo, and Alejandro Torres-Forné  for useful comments and helpful discussions. NSG acknowledges support from the Spanish Ministry of Science and Innovation via the Ram\'on y Cajal programme (grant RYC2022-037424-I), funded by MCIN/AEI/ 10.13039/501100011033 and by ``ESF Investing in your future”. This work is further supported by the Spanish Agencia Estatal de Investigaci\'on (Grant PID2021-125485NB-C21) funded by
MCIN/AEI/10.13039/501100011033 and ERDF A way
of making Europe,  
and by the European Union's Horizon 2020 research and innovation
(RISE) programme H2020-MSCA-RISE-2017 Grant No.
FunFiCO-777740 and by the European Horizon Europe staff exchange (SE) programme HORIZON-MSCA2021-SE-01 Grant No. NewFunFiCO-101086251. This work is also supported by the Multi-Annual Financing Program for R\&D Units,  2022.04560.PTDC (\url{https://doi.org/10.54499/2022.04560.PTDC}) and 2024.05617.CERN (\url{https://doi.org/10.54499/2024.05617.CERN}). The authors acknowledge the computer resources provided by the Red Espa\~nola de Supercomputaci\'on (Tirant, MareNostrum5, and Storage5) and the technical support from the IT departments of the Universitat de Val\`encia and the Barcelona Supercomputing Center (allocation grants RES-FI-2023-1-0023, RES-FI-2023-2-0002, RES-FI-2024-2-0012, and RES-FI-2024-3-0007).  

\bibliography{num-rel2}

\newpage

\appendix
\section{Appendix A: Fixed Mirror Radius}\label{sec:mirror}

\begin{figure}[h!]
\begin{center}
\includegraphics[width=0.48\textwidth]{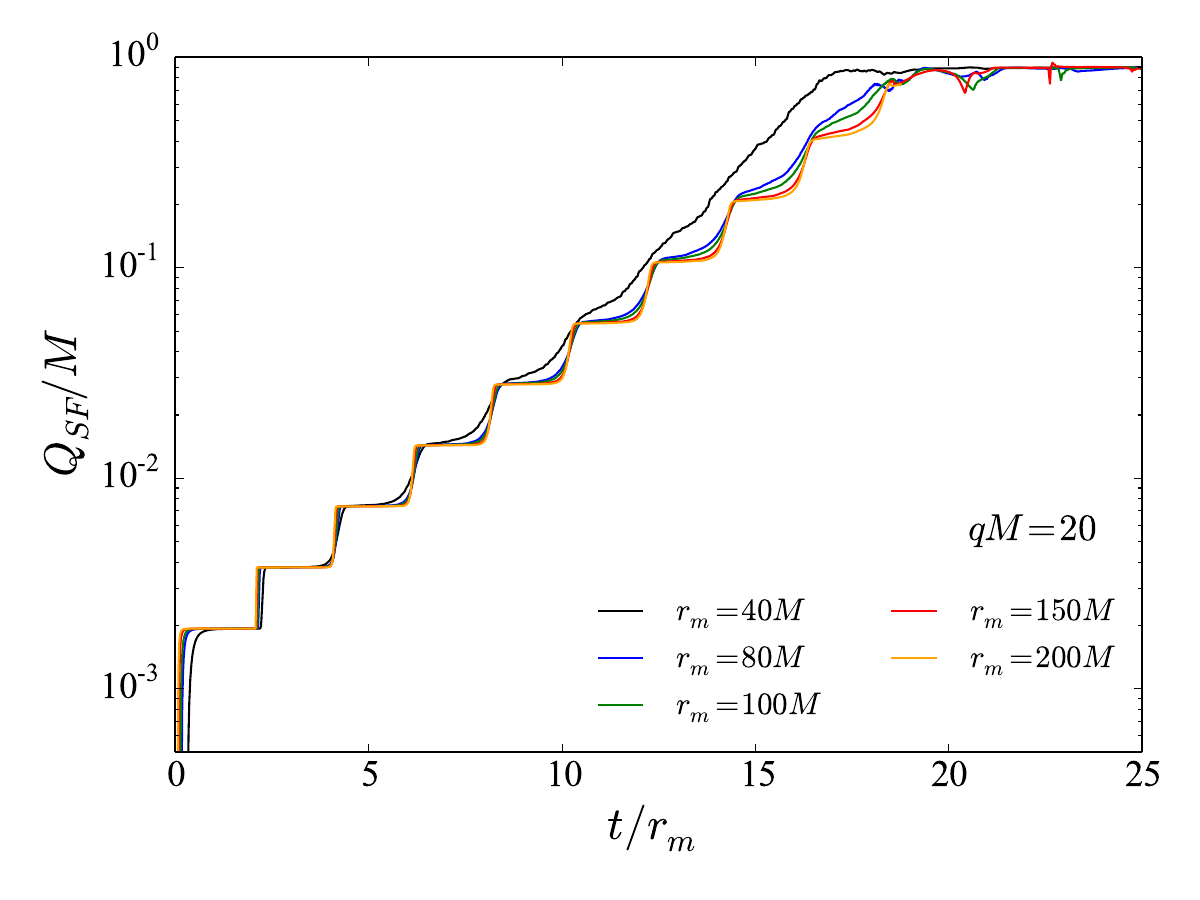}\\
\includegraphics[width=0.48\textwidth]{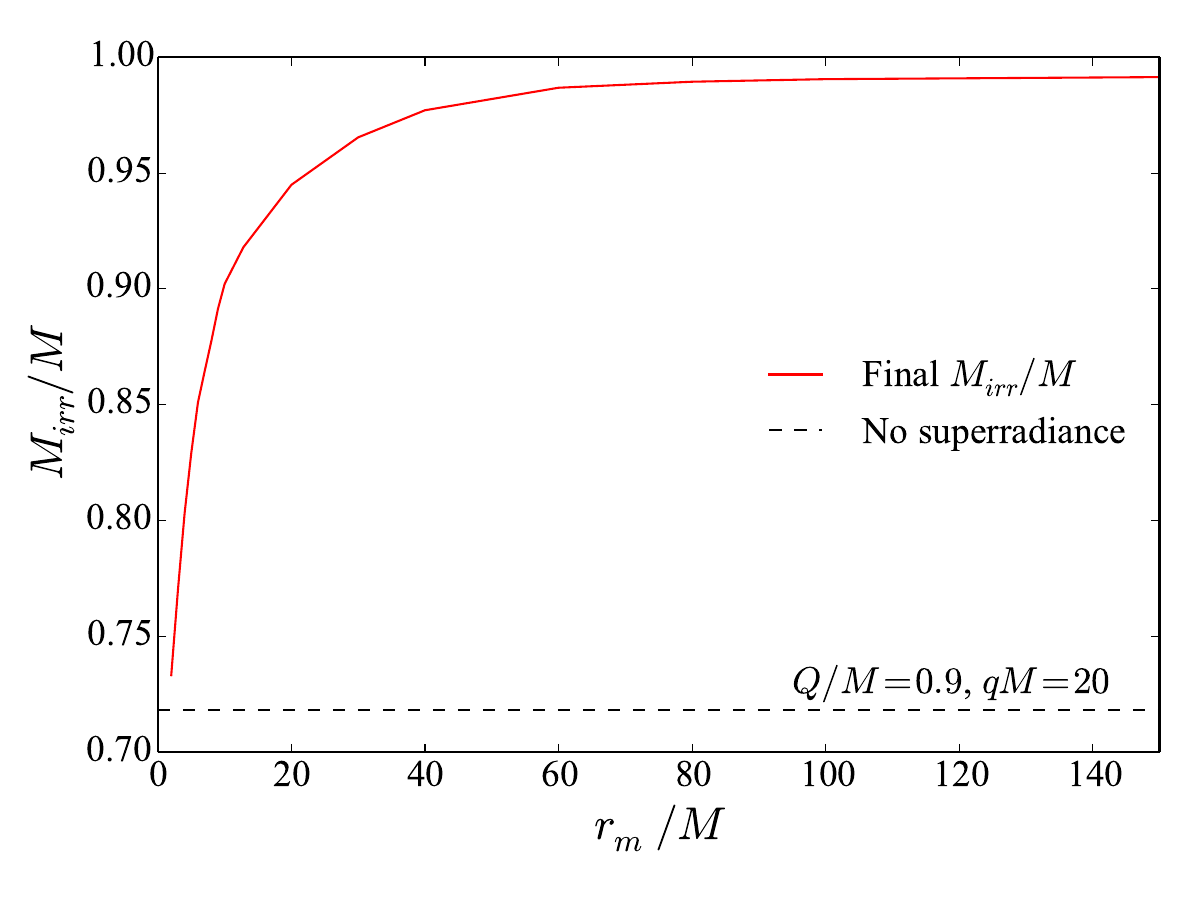}
\vspace{-0.5cm}\\
\caption{(Top panel) Time evolution of $Q_{\rm SF}$ for $qM=20$ configurations with different mirror radius $r_{m}$. Time has been rescaled with $r_{m}$ to highlight the scaling behavior of the instability. (Bottom panel) $M_{\rm irr}$ vs $r_m$  growth during the superradiant instability for $qM=20$. The final $M_{\rm irr}$ increases with $r_{m}$. The horizontal dashed black line corresponds to the initial $M_{\rm irr}/M$ for a $Q/M=0.9$ RN BH.}
\label{fig4}
\end{center}
\end{figure}

We explore how the instability timescale and the value of $M_{\rm irr}$ at the end of the superradiant instability change with the (fixed) mirror radius, for large values of $r_{m}$. In the top panel of Fig.~\ref{fig4} we show the time evolution of the scalar charge $Q_{\rm SF}$ during the superradiant phase. We find that the timescale increases linearly with $r_{m}$, indicating that for a mirror located at infinity the expected timescale would be infinite. In the bottom panel we plot the dependence of the final $M_{\rm irr}$ with respect to a fixed mirror with $r_m$ for $qM=20$. We observe that the irreducible mass of the final hairy black hole (BH) increases monotonically with the mirror radius.

\section{Appendix B: Hamiltonian constraint violations and convergence}

The Hamiltonian constraint can be used to monitor the accuracy of the simulation. It is defined as
\begin{equation}
    H\equiv R-(A^{2}_{a}+2A_{b}^{2})+\frac{2}{3}K^{2}-16\pi\mathcal{E}=0\,,
\end{equation}
where $K$ is the trace of the extrinsic curvature $K\equiv K^{i}_{i}$, $A_{a}$ and $A_{b}$ are the contraction of the traceless part of the conformal extrinsic curvature, 
$A_{a}\equiv\hat A^{r}_{r},\, A_{b}\equiv\hat A^{\theta}_{\theta}$ and $\mathcal{E}=n^{\mu}n^{\nu}T_{\mu\nu}$ is the energy density associated with the matter fields. Introducing reflective boundary conditions without specifying the stress-energy tensor of the mirror itself can lead to a constraint violations. Such constraint violations can result in non-physical evolutions~\cite{Okawa:2014nda}. In this section we will try to assess such limitations. 

\begin{figure}[h!]
\begin{center}
\includegraphics[width=0.445\textwidth]{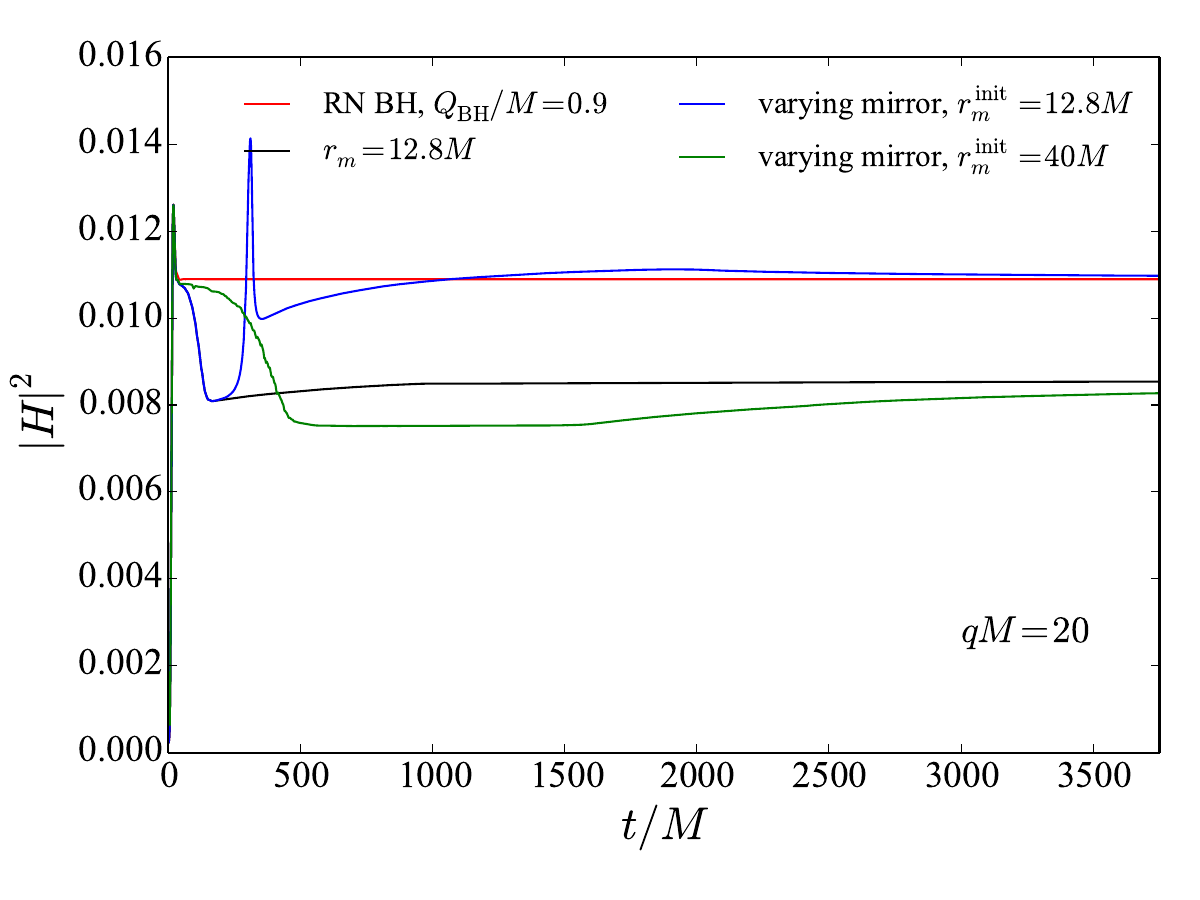}\\
\includegraphics[width=0.445\textwidth]{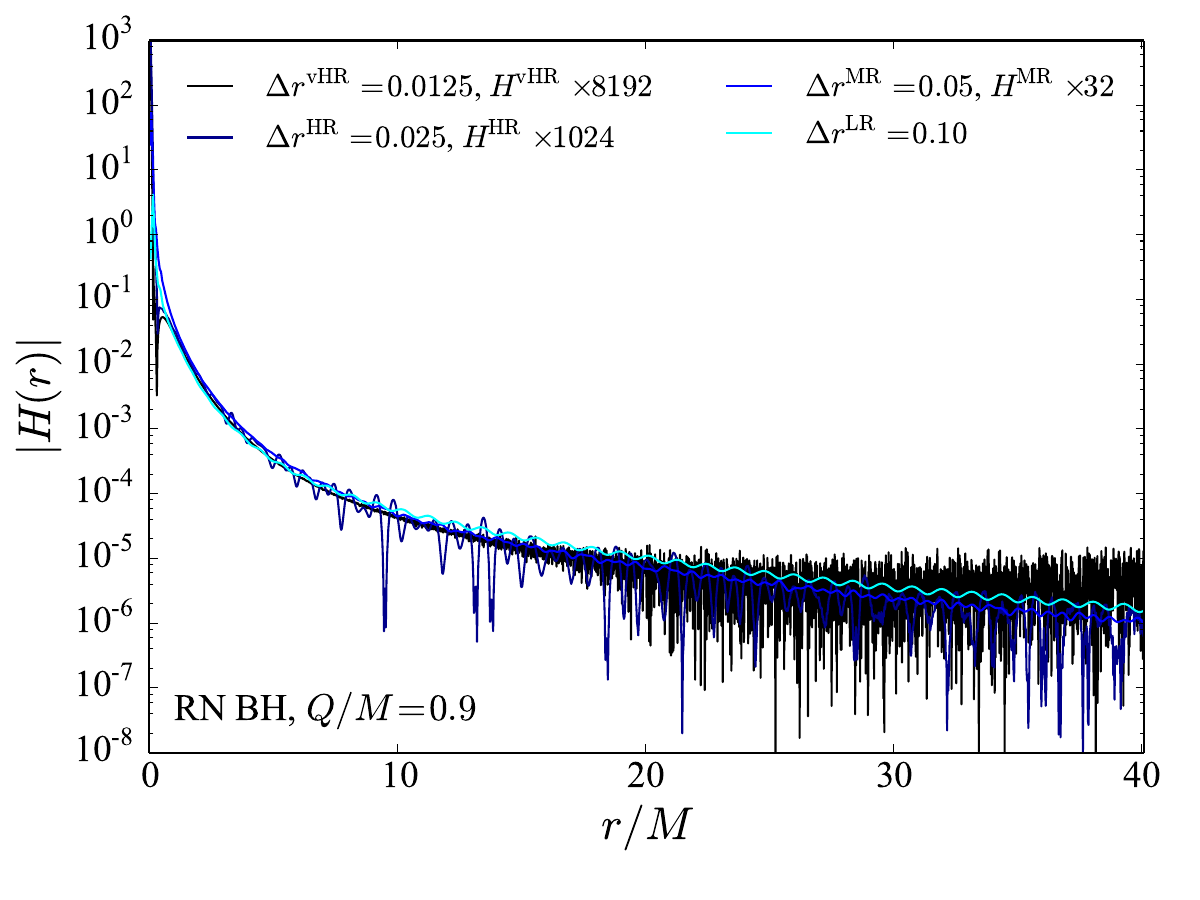}\\
\includegraphics[width=0.445\textwidth]{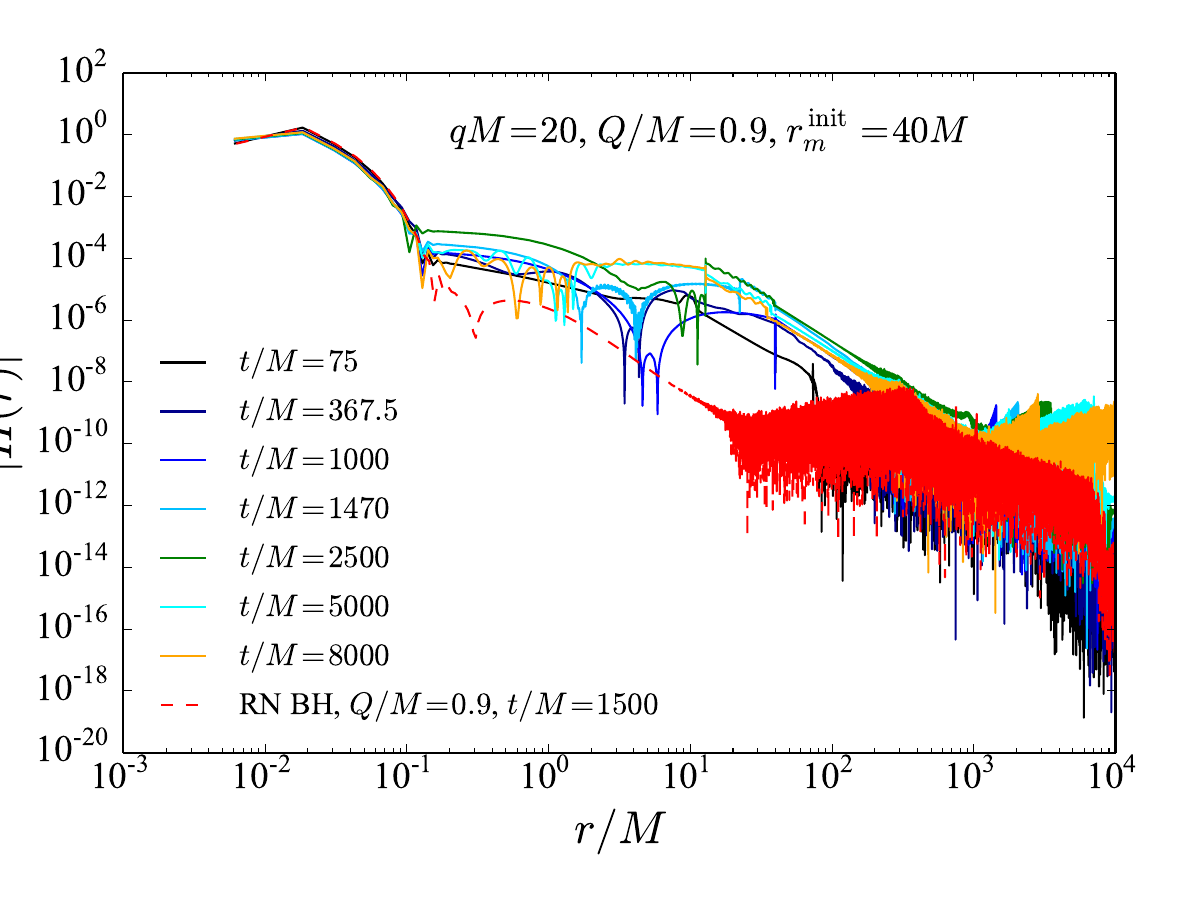}
\vspace{-0.5cm}\\
\caption{(Top panel): Time evolution of $\text{L}_2$ norm of the Hamiltonian constraint for the $qM=20$ configurations discussed in the main text (with $r^{\rm init}_m=12.8M$ and $v_m=-0.08$) and in Appendix D (with $r^{\rm init}_m=40M$, $r^{\rm final}_{m}=12.8M$, and $v_m=-0.08$). (Middle panel) Convergence test for a RN BH with $Q=0.9$ and $M=1$ at $t/M=1500$. The curves have been rescaled to approximately fourth-order convergence. (Bottom panel): Hamiltonian radial profile at different times for the configuration discussed in the Appendix D, compared to the RN BH case with the same resolution (red dashed line).} 
\label{fig5}
\end{center}
\end{figure}

\begin{figure}[h!]
\begin{center}
\includegraphics[width=0.45\textwidth]{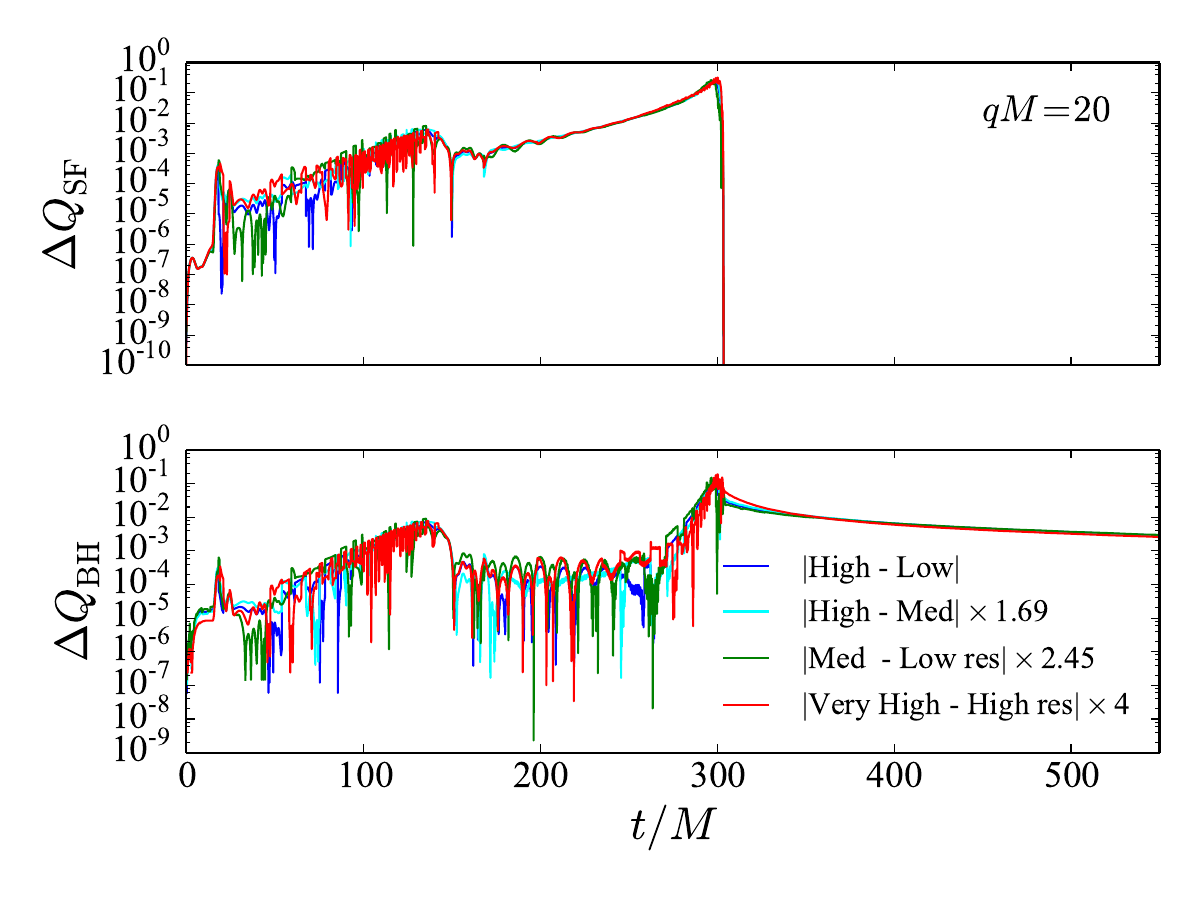}\\
\includegraphics[width=0.45\textwidth]{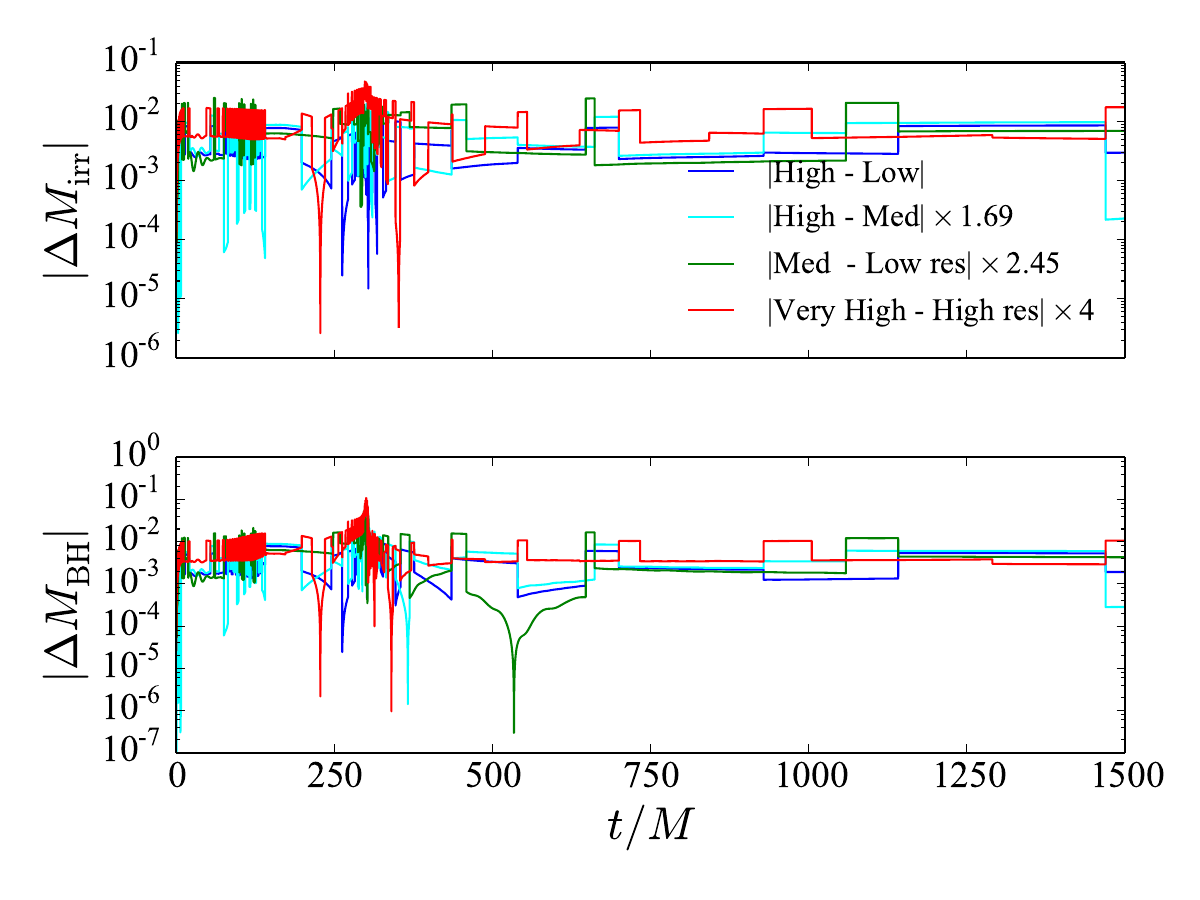}
\vspace{-0.5cm}\\
\caption{(Top panel): Convergence test for the configuration with $qM=20$ and $r^{\rm initi}=12.8M$ discussed in the main text, considering four different resolutions with $qM=20$: $\Delta r^{\rm vHR}=0.00625$ (very high resolution), $\Delta r^{\rm HR}=0.0125$ (high resolution), $\Delta r^{\rm MR}=5/3\Delta r^{\rm HR}$ (medium resolution), and $\Delta r^{\rm LR}=0.025$ (low resolution). We plot the difference between resolutions for $Q_{\rm SF}$ and $Q_{\rm BH}$ rescaled for second-order convergence. (Bottom panel): Convergence test for the total black hole mass $M_{\rm BH}$ computed from Christodoulou formula and the irreducible mass $M_{\rm irr}$ rescaled to second-order convergence. }
\label{fig6}
\end{center}
\end{figure}

In the top panel of Fig.~\ref{fig5} we plot the time evolution of the L2-norm of the Hamiltonian constraint for a Reissner-Nordström (RN) BH (red curve), for a superradiant configuration with $qM=20$ and a fixed mirror at $r_{m}=12.8M$ (black curve), and two setups with moving-mirror system with $qM=20$: one with $r^{\rm init}_m=12.8M$ and $r^{\rm final}=0$ (blue curve), and the other $r^{\rm init}_m=40M$ and $r^{\rm final}=12.8M$ (green curve). For the latter cases, the Hamiltonian constraint violations slightly increase as we move the mirror but once the scalar field is returned into the BH or the mirror reaches its final value, the value tends towards the corresponding value of the L2-norm. The middle panel displays the radial profile of the Hamiltonian constraint evaluated for a bald RN BH with $Q=0.9$ and $M=1$ at $t/M=1500$, for four different resolutions and rescale to fourth-order convergence. In the bottom panel we show the radial profile of the Hamiltonian constraint for the case with the mirror starting at $r_{m}^{\rm init}=40M$, $r_{m}^{\rm final}=12.8M$, $t_{\rm initial}=12500$ and $t_{\rm final}=1590$, discussed in Appendix D. The violations grow near the BH, with respect to the RN case (red dashed curve), but stay at low values ($\sim10^{-5}-10^{-4}$) during the evolution of the mirror. The puncture dominates the violations of the Hamiltonian constraint.  Finally, in Fig.~\ref{fig6} we show a convergence test taking four different resolution $\Delta r^{\rm vHR}=0.00625$ (very high resolution), $\Delta r^{\rm HR}=0.0125$ (high resolution), $\Delta r^{\rm MR}=5/3\Delta r^{\rm HR}$ (medium resolution), and $\Delta r^{\rm LR}=0.025$ (low resolution). We plot the difference between resolutions for $Q_{\rm SF}$, $Q_{\rm BH}$, the irreducible mass $M_{\rm irr}$, and the total BH $M_{\rm BH}$ mass computed from:
\begin{equation}
M_{\rm BH}^2=M_{\rm irr}^2+\frac{Q^2_{\rm BH}}{4M_{\rm irr}^2}.
\end{equation}
The curves are rescaled for the expected second-order convergence.

\section{Appendix C: Comparison with 3D simulations}

We have also studied the system with a moving mirror with 3D simulations using the open-source code \textsc{einstein toolkit}~\cite{ET_web, Loffler2012} within the \textsc{cactus} framework and employing mesh refinement. The left-hand side of the Einstein equations is solved using the \textsc{lean} code, based on the BSSN formulation, while Maxwell equations are solved using the code of \cite{Zilhao_2015, CODE_PS_web, witek_2023_7791842} adapted for a massless vector field coupled to the charged complex scalar field. Our setup assumes equatorial plane symmetry and symmetry with respect to the $x$-axis in a grid with ten refinement levels and the spatial (coordinate) domain of each level is given by $\{512.0, 128.0, 64.0, 32.0, 16.0, 16.0, 16.0, 1.5, 0.75, 0.5\}M$, with a corresponding spatial resolution for each level of $\{6.4, 3.2, 1.6, 0.8, 0.4, 0.2, 0.1, 0.05, 0.025, 0.0125\}M$.

\begin{figure}[h!]
\begin{center}
\includegraphics[width=0.48\textwidth]{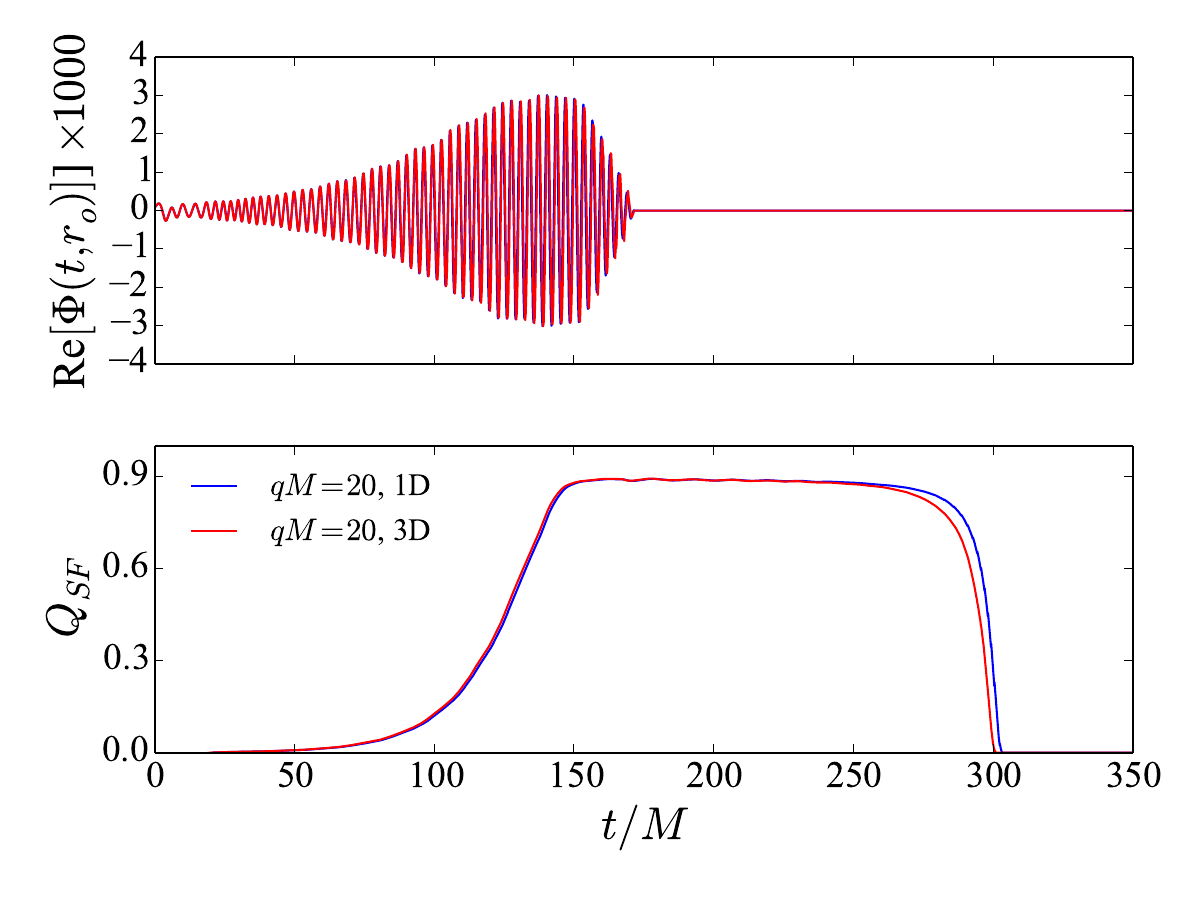}\\
\includegraphics[width=0.48\textwidth]{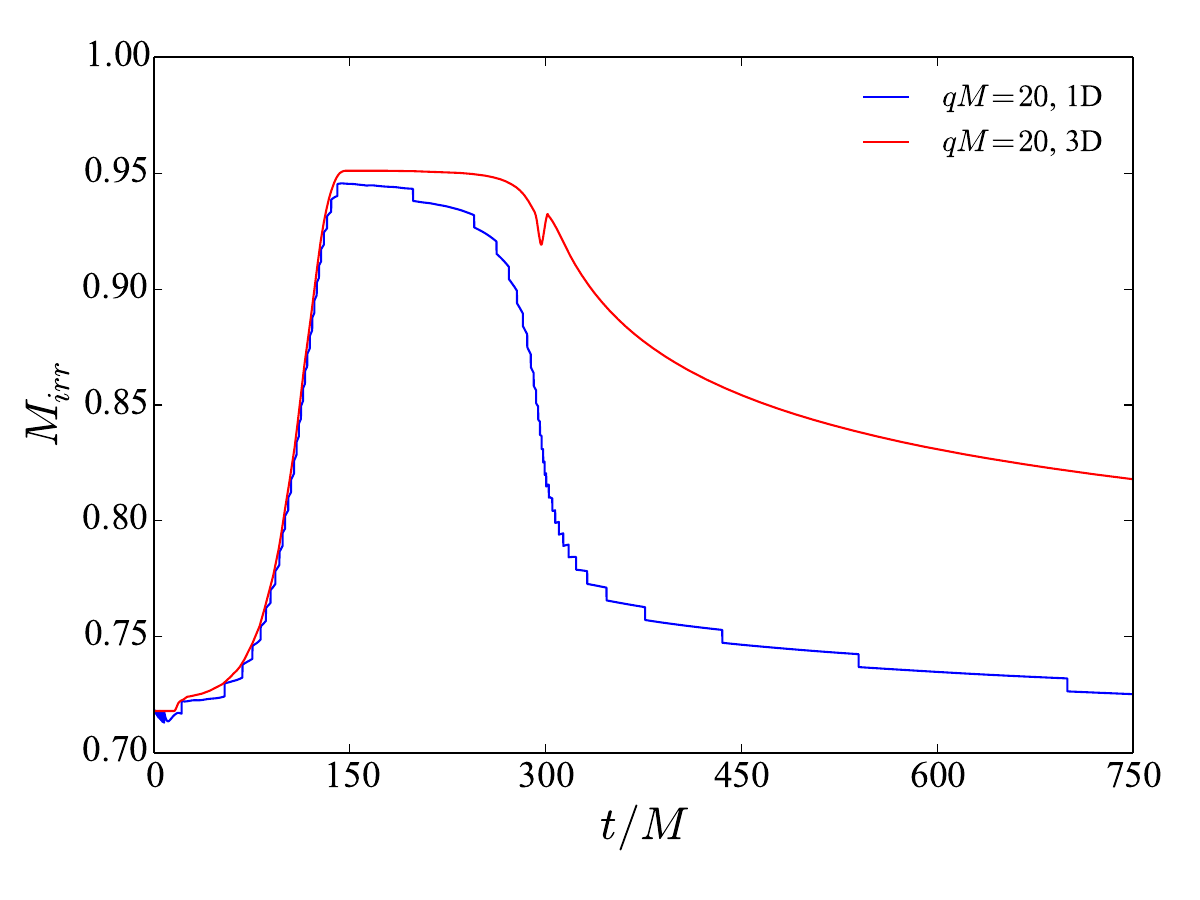}
\vspace{-0.5cm}\\
\caption{Comparison between 1D and 3D simulations. (Top panels) Time evolution of the real part of the scalar field extracted at $x=y=z=6.4M$ ($r_0=6.4\sqrt{2}M$ for $qM=20$ and $Q_{\rm BH}=0.9$ and the scalar field charge $Q_{\rm SF}$. (Bottom panel) Time evolution of the irreducible mass.}
\label{fig7}
\end{center}
\end{figure}

\begin{figure}[h!]
\begin{center}
\includegraphics[width=0.48\textwidth]{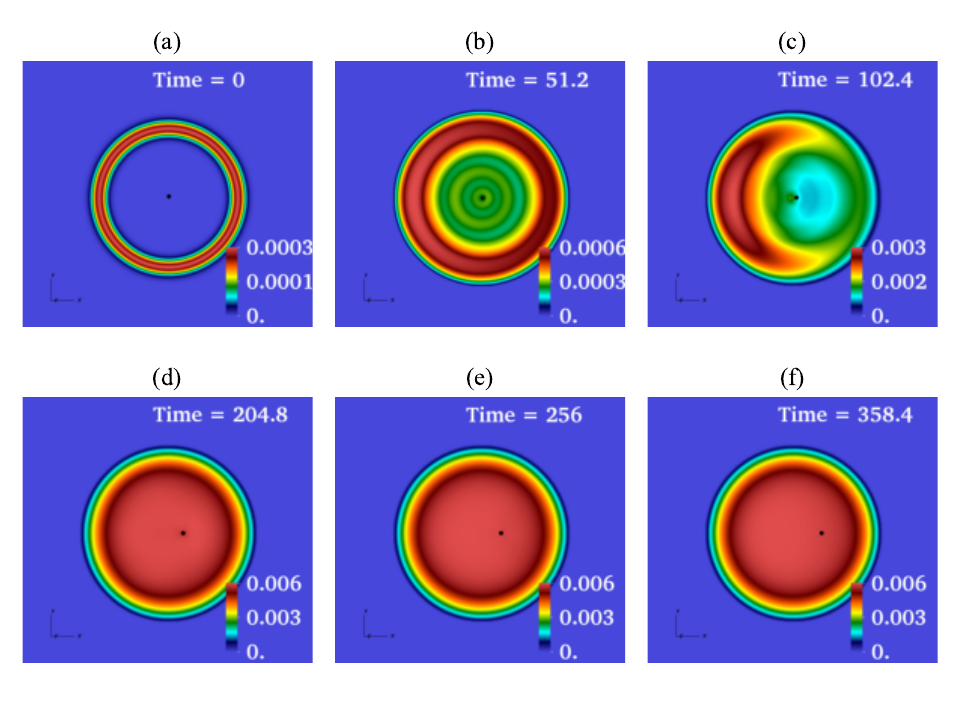}
\caption{Snapshots of the scalar field magnitude in the equatorial plane for an initial RN BH located at $x=10^{-3}M$ with $qM=20$ and $r_m=12.8M$ during the superradiant instability. The black dot corresponds to the position of the BH.}
\label{fig8}
\end{center}
\end{figure}

We have evolved the configuration described in case 2 with $qM=20$, $M=1$, $Q_{\rm BH}=0.9$, $r_m=12.8M$, $t_{\rm init}=150M$, $t_{\rm final}=310M$ and $v_m=-0.08$. In Fig.~\ref{fig7} we compare the time evolution of several illustrative quantities for the 1D and 3D evolutions. In particular in the top panels we show the amplitude of the real part of the scalar field as a function of time, extracted at $r_{0}=6.4\sqrt{2}M$ and the scalar field charge $Q_{\rm SF}$. As expected when the mirror radius becomes smaller than $r_{0}$, the scalar field goes to zero. The scalar field charge decreases more rapidly in the 3D case, likely due to the use of Cartesian coordinates, which results in the sphericity of the cavity getting worse resolved despite the refinement levels. This also affects the evolution of the apparent horizon mass. In the 3D case it also decreases and seems to tend to the initial value but at slower rate.

Furthermore, 3D evolutions allow us to explore the effect of non-spherical perturbations. In Fig.~\ref{fig8} we show the snapshots of the evolution of the scalar field magnitude with the initial BH placed at ($x=10^{-3}M$, $y=z=0$) instead of at the origin. We observe that the growth of the scalar field during the superradiant phase is not perfectly spherical leading to a recoil (kick) of the BH, which escapes the cavity. We conclude that even a small deviation from spherical symmetry can trigger an instability that disrupts the equilibrium configuration. This seems to fit the pattern that was unveiled for the mechanical instability of the BHs with synchronized/resonant hair, at least in the simplest models, as discussed in~\cite{Jordan2025}.

\section{Appendix D: Large mirror radii}
\begin{figure}[h!]
\begin{center}
\includegraphics[width=0.49\textwidth]{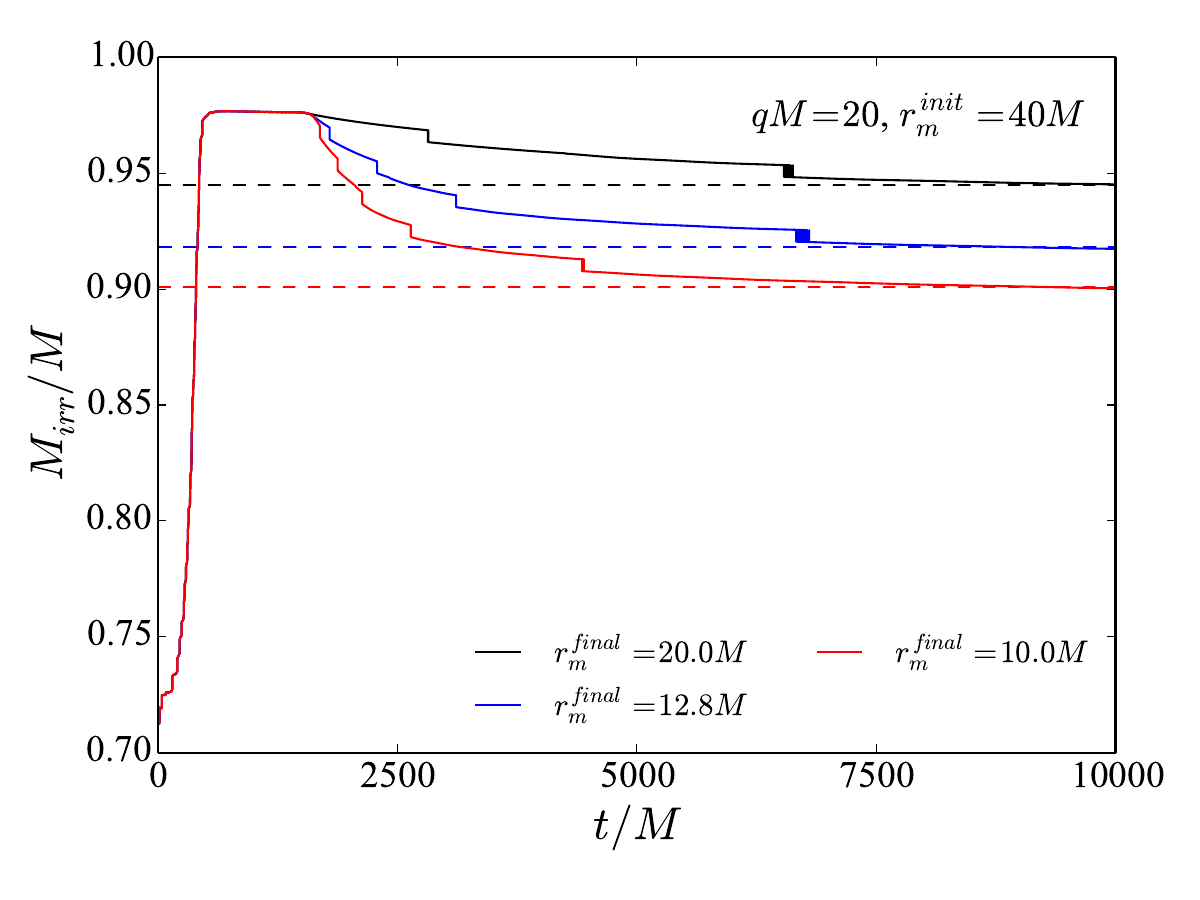}\\
\includegraphics[width=0.49\textwidth]{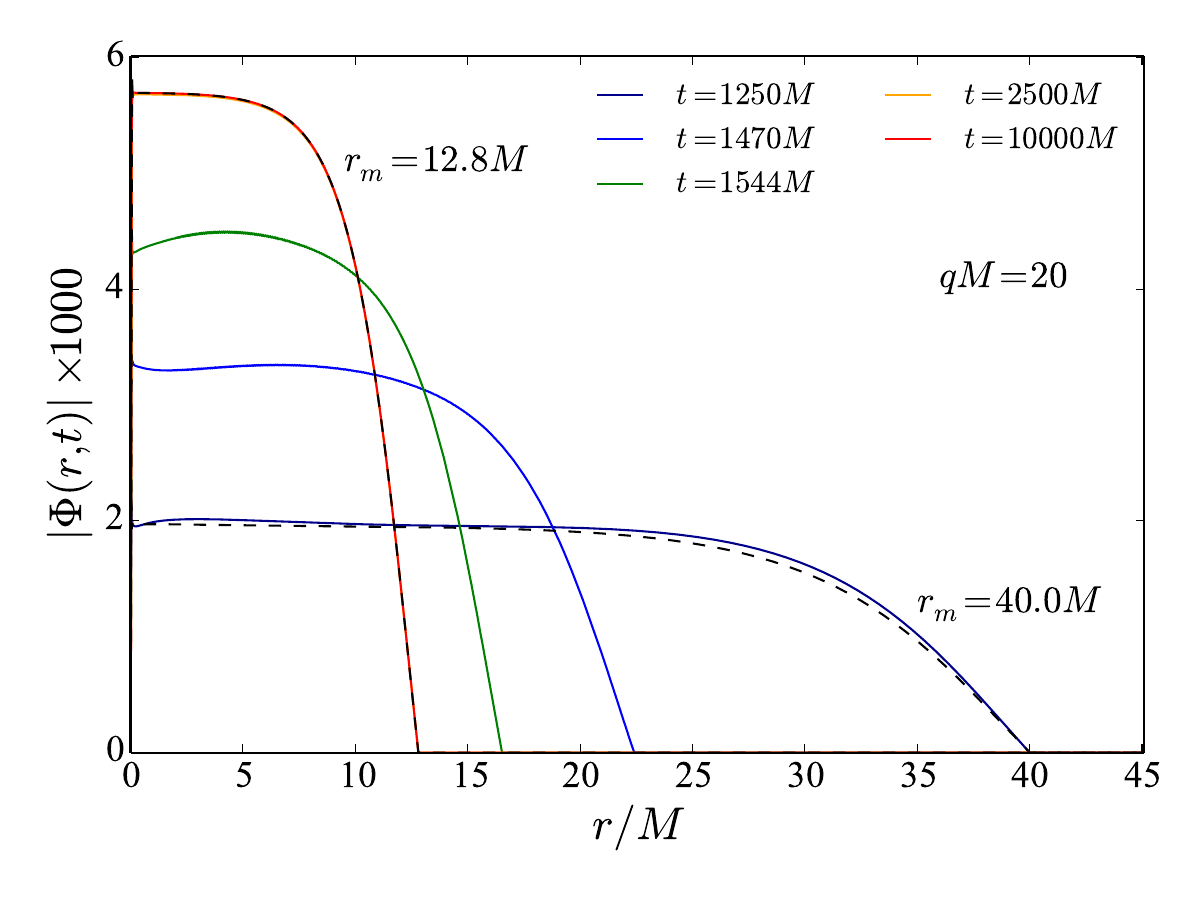}
\vspace{-0.5cm}\\
\caption{(Top panel) Time evolution of the irreducible mass for three different configurations with $qM=20$, $r^{\rm init}_m=40M$, $v_m=-0.08$ and $r^{\rm final}_m=\lbrace 10, 12.8, 20\rbrace M$. The horizontal dashed lines correspond to the final irreducible mass for the fixed mirror case. (Bottom panel) Radial profiles of the scalar field magnitude at different times for the shrinking mirror case with $qM=20$ and $r^{\rm final}_m= 12.8 M$.}
\label{fig9}
\end{center}
\end{figure}

We explore the behaviour of the irreducible mass in the case of larger initial mirror radii to avoid any potential nonphysical effect from imposing reflective boundary conditions too close to the horizon and to reinforce the robustness of our conclusions. We start with a mirror located at $r^{\rm init}_m=40M$ and evolve the system until $t^{\rm init}=1250M$, time at which the hairy BH configuration is almost in equilibrium. We then move the mirror with an inward motion and constant speed $v_m=-0.08$ until we reach three different final values, namely $r^{\rm final}_m=\lbrace 10, 12.8, 20\rbrace M$, in three different simulations. As shown in the top panel of Fig.~\ref{fig9}, we observe the same qualitative reduction of $M_{\rm irr}$, which at late times converges to the value of $M_{\rm irr}$ of the corresponding equilibrium configuration.  The bottom panel of Fig.~\ref{fig9} displays the radial profiles of the scalar field magnitude at different times. As the system evolves, the frequency of the field $\omega$ matches at all times the critical frequency given by $q\phi_{\rm H}$. The final distribution of the field (orange and red lines) matches that of fixed system with a mirror at the same radius. Furthermore the final scalar and BH charges coincide with those of the fixed equilibrium solution.

\section{Appendix E: Energy balance with a moving mirror}

The energy and charge exchange between the BH and the charged scalar field during superradiance can be understood from the first law of black hole thermodynamics:
\begin{equation}
    \delta M_{\rm BH}=\frac{\kappa}{8\pi}\delta A_{\rm AH}+\phi_{\rm H}\delta Q_{\rm BH},
\end{equation}
where $\kappa$ is the surface gravity and $\phi_{\rm H}$ is horizon potential. For a fixed mirror, superradiance results in:
\begin{eqnarray}
&&\delta M_{\rm BH} = -\delta E_{\rm SF}<0,\nonumber\\
&&\delta Q_{\rm BH} = -\delta Q_{\rm SF}<0.\nonumber
\end{eqnarray}
During this phase and until reaching the endpoint, BH mass and charge are directly transferred to the scalar field. The process is compatible with the second law of thermodynamics stating that the horizon area should increase (or remain constant if the process is adiabatic); in fact, considering that the loss of charge and mass of the BH is due to a scalar particle with, respectively, charge $q$ and frequency $\omega>0$, then:
\begin{equation}
\delta A_{\rm AH} = \frac{8\pi}{\kappa}\,\bigl(\underbrace{\delta M_{\rm BH}}_{\leq0}-\underbrace{\phi_{\rm H}\,\delta Q_{\rm BH}}_{<0}\bigl)=\frac{8\pi}{\kappa}\,\bigl(-\omega+|q\phi_H| \bigl)\ge0 \ ,
\end{equation}
which holds because of the superradiance condition $\omega<|q\phi_H|$. Physically, 
the area law holds because the loss of electric potential energy of the BH is necessarily larger or equal to the loss of mass, 
$|\phi_{\rm H}\delta Q_{\rm BH}|\geq|\delta M_{\rm BH}|$.

Now let us investigate the moving mirror scenario. We will consider a schematic, illustrative example to gain insight into the conceptual and qualitative behaviour of the system, extrapolating from our results (the system evolving from one equilibrium configuration to another). As in our simulations, we will assume perfect spherical symmetry and reflective boundary conditions on the scalar field, so there is no radiation. 

Consider that initially the mirror is located (as a limiting process) at infinity, $r_{m}\rightarrow\infty$, surrounding an RN BH. By the end of the subsequent superradiant instability,  all charge should have been extracted by the scalar field (in an infinity timescale), but the final BH retains \textit{all the mass}.  

\begin{figure}[h!]
\begin{center}
\includegraphics[width=0.49\textwidth]{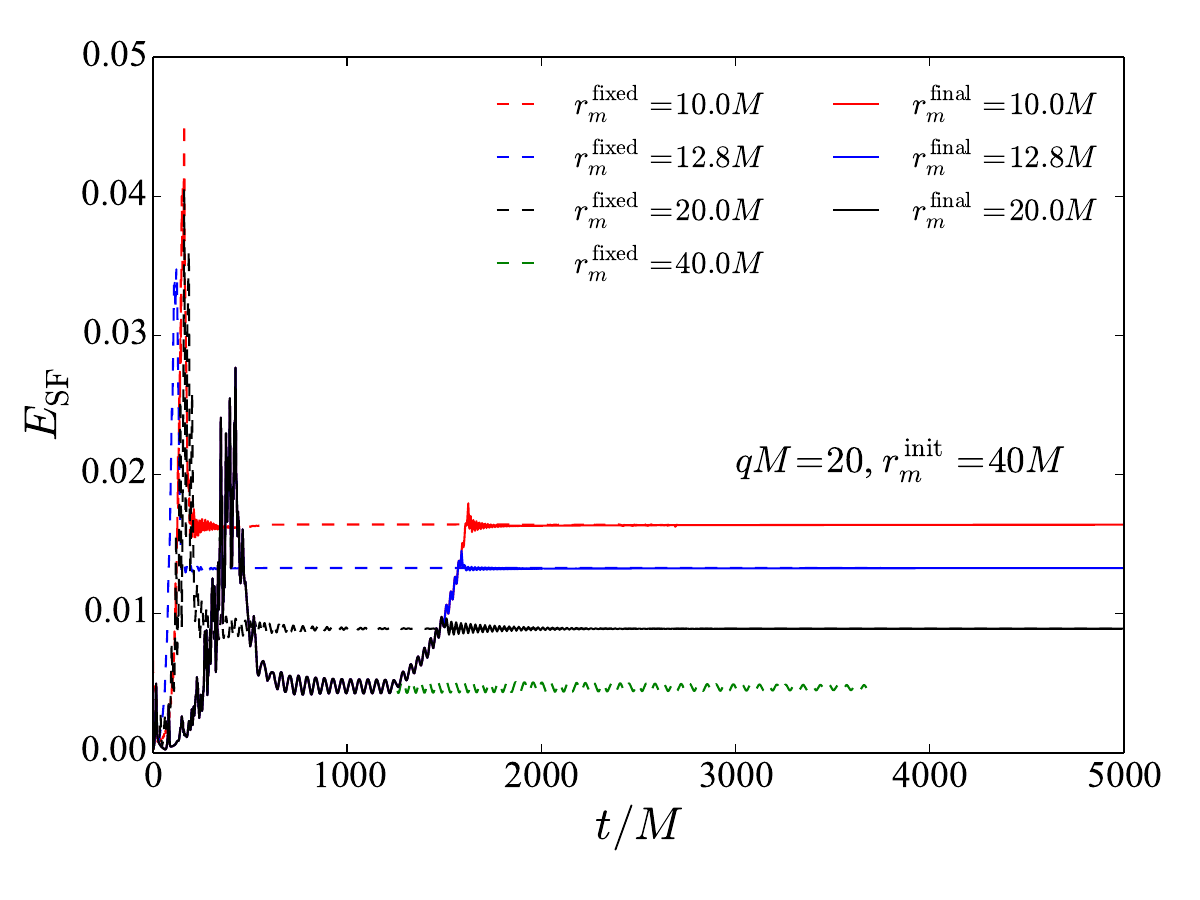}
\vspace{-0.5cm}\\
\caption{Same as Fig.~\ref{fig9} but for the scalar field energy $E_{\rm SF}$.  Time evolution of $E_{\rm SF}$ for three different configurations with $qM=20$, $r^{\rm init}_m=40M$, $v_m=-0.08$ and $r^{\rm final}_m=\lbrace 10, 12.8, 20\rbrace M$. The horizontal dashed lines correspond to the fixed mirror cases.}
\label{fig10}
\end{center}
\end{figure}

Thus, during the initial superradiant phase with a fixed mirror (at infinity), $\delta M_{\rm BH}=0$, and $\delta A_{\rm AH}=8\pi/\kappa\,\bigl(-\phi_{\rm H}\delta Q_{\rm BH}\bigl)>0$. Therefore, the transfer of charge to the scalar field implies that the area $\delta A_{\rm AH}>0$ and the irreducible mass $M_{\rm irr} \rightarrow M_{\rm BH}$, while the scalar field energy $E_{\rm SF}\rightarrow0$. 

On the other hand, during the subsequent phase of pushing the mirror inward $(\delta r_{m}<0$),   the BH gains charge, $\delta Q_{\rm BH}>0$, at the expense of the scalar cloud, $\delta Q_{\rm SF}<0$. If the mirror moves all the way to the horizon, the BH will absorb all the charge stored in the vanishing scalar field, $\delta Q_{\rm BH}=-\delta Q_{\rm SF}>0$ and $Q_{\rm BH}\rightarrow Q_{\rm SF}$. Comparing the initial (mirror at infinity) and final (mirror at $r_m=r_{\rm AH}$) states, the BH will not change its total mass $(\delta M_{\rm BH}=0)$; thus:
\begin{equation}
\delta A_{\rm AH} = \frac{8\pi}{\kappa}\,\bigl(-\underbrace{\phi_{\rm H}\,\delta Q_{\rm BH}}_{>0}\bigl)<0.
\end{equation}
This implies that the horizon area and the irreducible mass must decrease since the BH is forced to absorb more charge than mass, as seen in the evolutions where we take a shrinking cavity all the way to $r_m\rightarrow0$. 

Let us add the the behaviour of the scalar field energy with the radius is non-monotonic in this scenario. It starts with vanishing scalar field energy at $r_m\rightarrow \infty$, and then $E_{\rm SF}$ grows with decreasing $r_m$ down to a critical radius after (larger than the horizon radius), after which essentially all scalar field is driven into the BH.  The existence of a critical radius has been seen in perturbative studies of superradiance of a RN BH in a cavity \cite{Herdeiro:2013pia} where it was interpreted as due to the minimal wavelength (maximal frequency) allowed by the superradiance condition in this setup.

Finally, for a dynamical mirror moving from a finite radius to a smaller finite radius (larger than the critical radius), the energy of the scalar field increases, and thus the energy of the BH decreases, and as such, the horizon area (and hence the irreducible mass) should decrease:
\begin{equation}
\delta A_{\rm AH} = \frac{8\pi}{\kappa}\,\bigl(\underbrace{\delta M_{\rm BH}}_{<0}-\underbrace{\phi_{\rm H}\,\delta Q_{\rm BH}}_{>0}\bigl)<0.
\end{equation}
Indeed, our results show a clear increase of the scalar field energy as the cavity shrinks, $\delta E_{\rm SF}>0$ - see Fig.~\ref{fig10}, where we plot the scalar field energy computed as a volume integral of the field density.

\section{Appendix F: Schwinger pair production}
In our setup we have only considered classical fields which produce strong electric fields. However, when the field strength becomes sufficiently intense, quantum effects can no longer be neglected. In particular, the Schwinger effect~\cite{schwinger1951gauge} predicts that sufficiently strong electric fields can spontaneously produce electron–positron pairs from the vacuum. This results in a net loss of charge from the system. Pair production becomes significant when the electric field approaches or exceeds a critical value given by:
\begin{equation}
    E_{\rm crit}=\frac{m^{2}_{\rm e} c^{3}}{q_{\rm e}\hbar}\approx1.3\times10^{18}\,\,{\rm V/m},
\end{equation}
where $m_{\rm e}$ is the mass of the electron and $q_{\rm e}$ is the elementary charge. In natural units $c=G=\hbar=1$, this reduce to:
\begin{equation}
    E_{\rm crit}\sim\frac{m^{2}_{\rm e}}{q_{\rm e}}\sim10^{-6},
\end{equation}
and the electromagnetic invariant:
\begin{equation}
    F_{\mu\nu}F^{\mu\nu}\sim-2\times10^{-12}.
\end{equation}

\begin{figure}[h!]
\begin{center}
\includegraphics[width=0.49\textwidth]{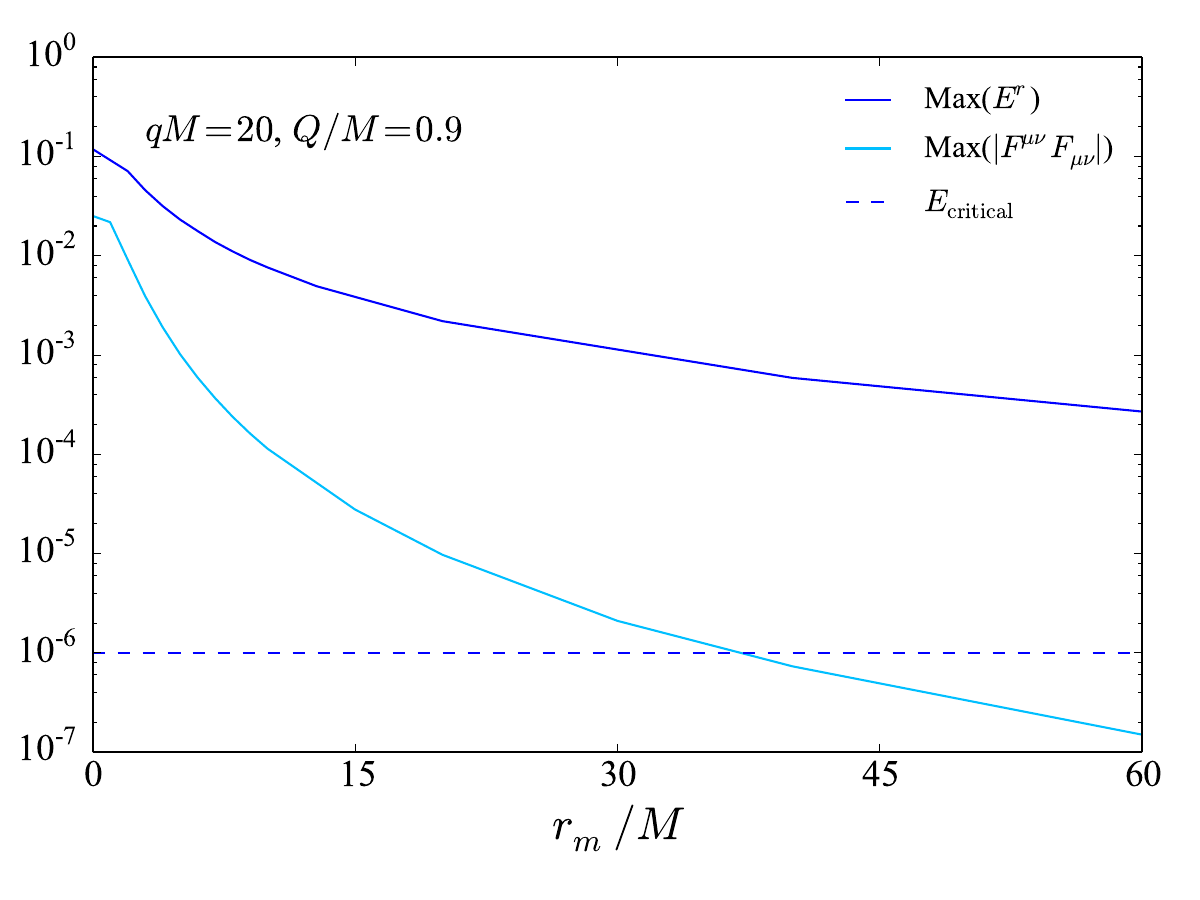}
\vspace{-0.5cm}\\
\caption{Maximum value of the $r-$component of the electric field $E^{r}$ and of the electromagnetic invariant $F_{\mu\nu}F^{\mu\nu}$ of hairy BH solutions for different values of the mirror radius $r_m$. The dashed line correspond to the critical Schwinger electric field strength.}
\label{fig11}
\end{center}
\end{figure}

We checked the maximum value of the electric field and of the electromagnetic invariant for our hairy BH solutions. In Fig.~\ref{fig11} we plot such quantities as a function of the mirror radius. We estimate that for mirror radii $r_m/M\lesssim1000$, the field strength would be larger than $E_{\rm critical}$, triggering Schwinger production and likely preventing the process described in this work.

\end{document}